\documentclass[aps,prb,showpacs,raggedbottom,nobalancelastpage,amssymb]{revtex4}
\usepackage{amsmath}
\usepackage{amsfonts}
\usepackage{graphicx}
\usepackage{appendix}
\usepackage{dsfont}
\usepackage{amssymb}
\usepackage{bm}
\usepackage{color}
\usepackage{ulem}
\usepackage{soul}
\usepackage{natbib}
\usepackage{nccmath}
\usepackage{array}
\newcommand{\beq}{\begin{equation}}
\newcommand{\eneq}{\end{equation}}
\newcommand{\be}{\begin{equation}}
\newcommand{\ee}{\end{equation}}
\newcommand{\bea}{\begin{eqnarray}}
\newcommand{\eea}{\end{eqnarray}}

\usepackage{multirow}
\usepackage{wasysym}

\begin{document}
\title{
Kondo length in bosonic lattices}
\author{Domenico Giuliano$^{(1,2)}$, Pasquale Sodano$^{(3,4)}$, 
and Andrea Trombettoni$^{(5,6)}$}
\affiliation{
$^{(1)}$ Dipartimento di Fisica, Universit\`a della Calabria Arcavacata di 
Rende I-87036, Cosenza, Italy \\
$^{(2)}$ I.N.F.N., Gruppo collegato di Cosenza, 
Arcavacata di Rende I-87036, Cosenza, Italy\\
$^{(3)}$International Institute of Physics, Universidade Federal 
do Rio Grande do Norte, 59078-400 Natal-RN, Brazil\\
$^{(4)}$Departemento de F\'isica Teorica e Experimental,
Universidade Federal do Rio Grande do Norte, 59072-970 Natal-RN, Brazil\\
$^{(5)}$ CNR-IOM DEMOCRITOS Simulation Center, Via Bonomea 265, 
I-34136 Trieste, Italy\\ 
$^{(6)}$ SISSA and INFN, Sezione di Trieste, Via Bonomea 265, I-34136 
Trieste, Italy}
\date{\today}

\begin{abstract}
Motivated by the fact that the low-energy properties of 
the Kondo model can be effectively simulated in spin chains, 
we study the realization of the effect with bond impurities 
in ultracold bosonic lattices at half-filling. 
After presenting a discussion of the effective theory and of the mapping 
of the bosonic chain onto a lattice spin Hamiltonian, we provide 
estimates for the Kondo length as a function 
of the parameters of the bosonic model. We point out that 
the Kondo length can be extracted from the integrated real space
correlation functions,  which are experimentally accessible 
quantities in experiments with cold atoms.   
\end{abstract}

\maketitle

\section{Introduction}
\label{intro}

The Kondo effect has been initially studied in metals, like 
Cu, containing magnetic impurities, like Co atoms, where it arises 
from the interaction between magnetic impurities and conduction electrons, 
resulting in a net, 
low-temperature increase of the resistance \cite{kondo64,hewson,kouwenhoven01}. It soon assumed a prominent role 
in the description of strongly correlated systems and 
in motivating and benchmarking the development of 
(experimental and theoretical) tools to study them \cite{hewson,Wilson1975}.
Indeed, due to the large amount of analytical and numerical 
tools developed to attack it, the Kondo effect has  become a paradigmatic example of a strongly interacting system and a 
testing ground for a number of different  many-body techniques.
 
 The interest in the Kondo effect significantly revitalized when it became 
possible to realize  it in a controlled way in a solid 
state system, by using quantum dots in contacts with metallic 
leads, in which the electrons 
trapped within the dot can give rise to a net nonzero total spin interacting 
with the spin of conduction electrons from the leads, 
thus mimicking the behavior of a magnetic impurity in a metallic host 
\cite{alivisatos96,kouwenhoven98,gg98_1,gg98_2}. An alternative 
realization of Kondo physics is recovered within the universal, 
low energy-long distance 
physics of a magnetic impurity  coupled to a gapless antiferromagnetic chain \cite{sorensen,affleck08}. 
  In fact, though low-energy excitations of a spin chain are realized as collective spin
modes, the remarkable phenomenon of ''spin fractionalization'' \cite{fadeev} implies that the actual stable 
elementary excitation of an antiferromagnetic spin-1/2 spin chain is a spin-1/2 ''half spin wave'' \cite{haldane,giulaugh_1} (dubbed 
spinon). Spinons have a gapless spectrum and, therefore, for what concerns screening of the impurity spin, 
they  act exactly as itinerant electrons in metals, as the charge quantum number is completely irrelevant 
for Kondo physics.  
A noticeable  advantage of working with the  spin chain realization of the Kondo effect   
is that a series of tools developed for spin systems, including entanglement witnesses and negativity, can be used 
to study the Kondo physics in these systems  \cite{bayat10,bayat12}.

Another important, 
long-lasting reason for interest in Kondo systems lies in that
the multichannel ''overscreened'' version of the effect \cite{noz_bla,AFFLECK_LUDWIG} 
provides a remarkable realization of non-Fermi liquid behavior  \cite{Nandrei}. Finally,  
the nontrivial properties of Kondo lattices  provide  
a major arena in which to study many-body nonperturbative effects,   
related to heavy-fermion materials \cite{lattice_1,lattice_2}. 
A recent example of both theoretical and experimental activity on multichannel 
Kondo systems is provided by the topological Kondo model 
\cite{beri12,altland14,buccheri15,eriksson}, based on the merging of 
several one-dimensional quantum wires with suitably induced and 
possibly controllable Majorana modes tunnel-coupled at their edges,
and by recent proposals of realizing topological Kondo Hamiltonians 
  in $Y$-junctions of XX and Ising chains 
\cite{crampe13,tsvelik13,giu_stt} and of Tonks-Girardeau gases 
\cite{buccheri16}. Finally, the effects of the competition between
the Kondo screening and the screening from localized Majorana modes 
emerging at the interface between a topological superconductor and a 
normal metal has been recently discussed in [\onlinecite{giuaf_3}] using
the techniques developed in [\onlinecite{giuliano_affleck_1}].

The onset of the Kondo  effect is set by the  Kondo temperature $T_K$, 
which emerges from the perturbative renormalization group (RG)  approach 
as a scale at which the system crosses over towards the strongly correlated nonpertubative regime 
\cite{hewson,Wilson75}. The systematic implementation of RG techniques has 
clearly evidenced the scaling behavior characterizing the Kondo regime, which 
results in the collapse onto each other  of the curves describing physical quantities in terms of 
the temperature $T$, once $T$ is rescaled by $T_K$ \cite{anderson_1,Wilson75}.   The 
collapse evidences the one-parameter scaling, that is, there is only one dimensionful quantity, 
which is dynamically generated by the Kondo interaction and invariant under RG trajectories. Thus, 
within scaling regime, one may trade $T$ for another dimensionful scaling parameter such
as, for instance, the system size $\ell$. In this case,  as a consequence of 
one-parameter scaling,  a scale invariant quantity with the 
dimension of a length emerges, the  Kondo screening length $\xi_K$, given by 
$\xi_K=\hbar v_F/k_B T_K$, where $v_F$ is the Fermi velocity of conduction 
electrons and $k_B$ is the Boltzmann constant \cite{Wilson75}. Physically, 
$\xi_K$ defines the length scale over which the impurity magnetic moment is fully screened 
by the spin of conduction electrons, that is, the ''size of the Kondo cloud'' \cite{Affleck09}.
 Differently from $T_K$, which can be directly measured from the low-$T$ behavior of the 
resistance in metals, the emergence of $\xi_K$ has been so far only theoretically 
predicted, as a consequence of the onset of the Kondo scaling \cite{Wilson75}. Thus, 
it would be extremely important to directly probe $\xi_K$, as an ultimate consistency check of
scaling in the Kondo regime.   As the emergence of the Kondo screening length is a mere 
consequence of the onset of Kondo scaling regime, $\xi_K$ can readily be defined for Kondo effect 
in spin chains, as well \cite{affleck_length,sorensen,bayat12}. Unfortunately,  
despite the remarkable efforts paied in the last years to estimate $\xi_K$ in various systems
by using combinations of perturbative, as well as nonperturbative
numerical methods  \cite{affleck08}, the Kondo length  still appears quite an elusive 
quantity to directly detect, both in solid-state electronic systems as well as 
in spin chains \cite{Affleck09}. This makes it desirable to investigate 
alternative systems in which to get an easier experimental access
to $\xi_K$.  

A promising route in this direction may be 
provided by the versatility in the control and manipulation of ultracold atoms 
\cite{pethick02,pitaevskii03}. Indeed, in the last years several proposals 
of schemes in which features of the Kondo effect can be studied 
in these systems have been discussed. 
Refs.[\onlinecite{recati02,porras08}] suggest to realize the spin-boson model 
using two hyperfine levels of a bosonic gas \cite{recati02}, or 
trapped ions arranged in Coulomb crystals \cite{porras08} (notice that 
in general the Kondo problem may be thought of as a spin-1/2,
system interacting with a fermionic bath \cite{leggett87}). 
Ref.[\onlinecite{duan04}] proposes to use ultracold atoms in 
multi-band optical lattices controlled through spatially periodic 
Raman pulses to investigate a class of strongly correlated physical systems  
related to the Kondo problem. Other schemes involve 
the use of ultracold fermions near a Feshbach resonance \cite{falco04}, 
or in superlattices \cite{paredes05}. More recently, 
the implementation of a Fermi sea of spinless fermions \cite{noshida13} or 
of two different hyperfine states of one atom species \cite{bauer13} 
interacting with an impurity atom of different species
confined by an isotropic potential has been proposed\cite{noshida13}. 
The simulation of the $SU(6)$ Coqblin-Schrieffer model for an ultracold 
fermionic gas of Yb atoms with metastable states has been discussed, while  
alkaline-earth fermions with two orbitals were also at the heart of the recent 
proposal of simulating Kondo physics through a suitable application 
of laser excitations \cite{nakagawa15}. Despite such an intense 
theoretical activity, including the investigation 
of optical Feshbach resonances to engineer Kondo-type 
spin-dependent interactions in Li-Rb mixtures \cite{sundar16}, 
and the remarkable progress in the manipulation of ultracold atomic systems, 
such as alkaline-earth gases, up to now an experimental detection 
of features of Kondo physics and in particular of the Kondo length 
in ultracold atomic systems is still lacking.

In view of the observation that optical lattices provide an highly controllable 
setup in which it is possible to vary the parameters of the Hamiltonian and 
to accordingly add impurities with controllable parameters\cite{morsch06,bloch08}, 
in this paper we propose to study the Kondo length in 
ultracold atoms loaded on an optical lattice. 
Our scheme is based on the well-known mapping between  the lattice Bose-Hubbard (BH) 
Hamiltonian and the XXZ spin-$1/2$ Hamiltonian 
\cite{matsubara56}, as well as on the Jordan-Wigner (JW)
representation for the spin $1/2$ operators, which allows 
for a further mapping onto   a Luttinger liquid model \cite{gogolin98,schulz00,giamarchi04}.
Kondo effect in Heisenberg spin-1/2 antiferromagnetic spin chains has been 
extensively studied  \cite{eggert92,furusaki98,sirker08}, though mostly for 
side-coupled impurities (i.e., at the edge of the chain). 
For instance,  in Ref.[\onlinecite{furusaki98}], the Kondo impurity 
is coupled to a single site of a gapless XXZ spin chain, while in Ref.[\onlinecite{sorensen}] a magnetic impurity 
is coupled at the end of a $J_1-J_2$ spin-$1/2$ chain. 
At variance,  in trapped ultracold atomic systems, it is usually 
difficult to create an impurity at the edge of the system. 
Accordingly,  in this paper we propose to study the Kondo length at 
an extended (at least two links) impurity realized in the bulk of a cold atom system on a   
$1d$ optical lattice. In particular,  we assume the lattice to be   at half-odd filling, so to avoid the 
onset of a gapped phase that takes place at integer filling in the limit
of a strong repulsive interaction between the particles.    
Since the real space correlation functions are quantities that one can measure 
in a real cold atom experiment, we address the issue of how to extract the Kondo length 
from the zeroes of the integrated real space density-density correlators.    
Finally, we provide estimates for $\xi_K$  and show that, for typical values of 
the system parameters, it takes values within the reach of 
experimental detectability   ($\sim$ tens of lattice sites).

  Besides the possible technical advances, we argue that, at variance with what 
happens   at a magnetic impurity in a conducting metallic host, where one measures 
 $T_K$ and infers the  existence of  $\xi_K$    from the   applicability of one-parameter scaling to the Kondo regime,
 in an ultracold atom setup one can extract from density-density correlation functions the Kondo screening length,
 that is in principle easier to measure, so that,
to access $\xi_K$, one has not to rely on verifying the one parameter scaling, which is what 
tipically makes $\xi_K$ quite hard to detect.

The paper is organized as follows:

\begin{itemize}
 \item In section \ref{mapping} we provide the effective description of 
 a system of ultacold atoms on a $1d$ optical lattice  as a spin-1/2 spin chain. In particular, 
 we show how to model impurities in the lattice corresponding to bond impurities in the spin chain;

\item In section \ref{ren_group} we derive the scaling equations for the Kondo running couplings and 
use them to estimate the corresponding Kondo length;

\item In section \ref{density_density} we discuss how to numerically extract the Kondo length from the integrated
real space density-density correlations and compare the results with the ones obtained in 
section  \ref{ren_group};

\item In section  \ref{conclusions} we summarize and discuss our results.
\end{itemize}
Mathematical details of the derivation and reviews of known results in the literature are
provided in the various appendices.

\section{Effective model Hamiltonian}
\label{mapping}

Based on the  spin-1/2 $XXZ$ spin-chain 
Hamiltonian description of (homogeneous, as well as inhomogeneous)
interacting bosonic ultracold atoms at half-filling in a deep 
optical lattice, in this section we propose  to model impurities in the spin chain by 
locally modifying  the strength of the link parameters of 
the optical lattice, eventually resorting to a model 
describing   two $XXZ$  ``half-spin chains'', 
interacting with each other via a local 
impurity. When the impurity is realized  as a spin-1/2 local spin, 
such a system corresponds to a possible realization 
of the (two channel) Kondo effect
in spin chains \cite{furusaki98,sorensen}. Therefore, our mapping leads to 
the conclusion that spin chain  Kondo effect may 
possibly realized and detected within bosonic cold
atoms loaded onto a one-dimensional optical lattice. 

 To resort to the spin-chain description of interacting ultracold atoms, we 
consider the  large on-site interaction energy $U$-limit of a system of  interacting ultracold bosons on  a deep 
one-dimensional lattice.  This is  described by the extended 
BH Hamiltonian \cite{fisher89,jaksch98,sampera12}
\beq
H_{\rm BH} = - \sum_{ j = - \ell}^{\ell - 1} t_{j;j+1} ( b_j^\dagger b_{j+1} 
+ b_{j+1}^\dagger b_j ) + \frac{U}{2} \sum_{j = - \ell}^\ell n_j ( n_j - 1 )
+ V \sum_{ j = - \ell}^{\ell - 1} n_j n_{j+1} - \mu  \sum_{j = - \ell}^\ell
n_j
\:\:\:\: .
\label{bhh1}
\eneq
\noindent
In Eq.(\ref{bhh1}), $b_j , b_j^\dagger$ 
are respectively the annihilation and the 
creation operator of a single boson at site $j$ (with 
$j= - \ell,\cdots,\ell$) and, accordingly, they 
satisfy the commutator algebra $ [ b_j , b_{j'}^\dagger ] = \delta_{ j , j'}$, 
all the other commutators being equal to $0$. As usual, we set 
$n_j= b_j^\dagger b_j$. 
Moreover, $t_{j;j+1}$ is the hopping amplitude for bosons between 
nearest neighboring sites $j$ and $j+1$, 
$U$ is the interaction energy between particles on the same site, $V$ is
the interaction energy between particles on nearest-neighboring sites.
Typically, for alkali metal atomes one has  $V \ll U$ while, for dipolar gases 
\cite{lahaye07} on a lattice, $V$ may be of the same order as $U$ 
\cite{altman02,dallatorre06}. 
Throughout all the paper we take $U>0$ and $V\geq 0$.   To outline the mapping onto 
a spin chain, we start by assuming that  $t_{j;j+1}$ is uniform across the 
chain and equal to $t$. Then, we discuss how to realize 
an impurity in the chain by means 
of a pertinent modulation of the $t_{j;j+1}$'s in real space.  
In performing the calculations, we 
will be assuming  open boundary conditions on the $2 \ell + 1$-site chain and  we will
set the average number of particles per site by fixing  the filling 
$f=\frac{N_T}{{\cal N}}$
where $N_T$ is the total number of particles 
on the lattice and ${\cal N} = 2 \ell + 1$ is the number of sites.

 In the large-$U$ limit, one may set up a mapping between the 
BH Hamiltonian in Eq.(\ref{bhh1}) and a pertinent 
spin-model Hamiltonian $H_S$, with $H_S$ either describing an integer, 
\cite{schulz86,dallatorre06}, or 
an half-odd spin chain \cite{grst_13}, 
depending on the value of $f$. An integer-spin effective Hamiltonian 
is recovered,   at large $U$,  for 
$f=n$ (with $n=1,2,\ldots$), corresponding to  
$\mu = \mu_0 ( n ) = n ( U + 2 V ) - U / 2$  and  $U \gg t$  \cite{altman02}, which 
allowed for recovering the phase diagram of the BH model in this limit by relating  
on the analysis of the phase diagram of   spin-$1$ chains 
within  the standard bosonization approach \cite{haldane83,schulz86}. In particular,  
the occurrence of Mott and Haldane gapped insulating phases for ultracold atoms 
on a lattice has been predicted and discussed 
\cite{dallatorre06,berg08,amico08}.   

Here, we rather focus onto the mapping of the BH Hamiltonian onto an effective spin-1/2 spin-chain 
Hamiltonian. This is recovered   at $U / t \gg 1$ and half-odd
filling $f = n + 1/2$ (with $n=0,1,2,\ldots$), corresponding to setting
the chemical potential so that 
$\mu = ( U + 2 V ) ( n + \frac{1}{2} ) - \frac{U}{2}$. In this regime, 
the effective low-energy spin-1/2 Hamiltonian for the system is given by
\cite{grst_13}

\beq
H_{\rm spin-1/2} = - J \sum_{ j = - \ell}^{\ell - 1} \left( 
S_j^+ S_{j+1}^- 
+ S_{j+1}^+ S_j^- \right) + 
J \Delta  \sum_{ j = - \ell}^{\ell - 1} S_j^z S_{j+1}^z
\:\:\:\: ,
\label{sp121}
\eneq
\noindent
  with the spin-1/2 operators $S_j^a$ defined as 
\begin{eqnarray}
 S_j^+ &=& \frac{1}{\sqrt{n + \frac{1}{2}}} \: {\bf P}_{\bf \frac{1}{2}}
 b_j^\dagger {\bf P}_{\bf \frac{1}{2}} \:\:\:\: , \nonumber \\
  S_j^- &=& \frac{1}{\sqrt{n + \frac{1}{2}}} \: {\bf P}_{\bf \frac{1}{2}}
 b_j  {\bf P}_{\bf \frac{1}{2}} \nonumber \:\:\:\: ,   \\
   S_j^z &=&  {\bf P}_{\bf \frac{1}{2}} [  b_j^\dagger b_j - f ]  {\bf P}_{\bf \frac{1}{2}}
\:\:\:\: .
\label{spinoperators}
\end{eqnarray}
\noindent
and  ${\bf P}_{\bf 1 /2}$ being the projector onto   the subspace of the Hilbert space 
${\cal F}_{\bf \frac{1}{2}}$, spanned by the states 
$\otimes_{ j = 1}^{\cal N} \left| n + \frac{1}{2} + \sigma \right\rangle_j$, with 
$\sigma = \pm \frac{1}{2}$. The parameters $J$ and $\Delta$  are  given by 
$J = \tilde{J} \left[ 1 - \frac{2 \tilde{J}}{U} \rho \right]$ and 
 $\Delta = \frac{\tilde{\Delta}}{\left[ 1 - \frac{2 \tilde{J}}{U} \rho \right] }$, 
with $ \tilde{J}= 2 t \left( n + \frac{1}{2} \right)$, $\tilde{\Delta} = \frac{V}{\tilde{J}} - 
\frac{ t^2 ( 2 n^2 + 6 n + 4 ) }{\tilde{J} U}
-  \frac{4  t^2 (  n + 1 )^2 }{\tilde{J} U}$ and  
$\rho = \frac{U ( n + 1 )}{2 \tilde{J}} - \sqrt{\left[ 
\frac{U ( n + 1 )}{2 \tilde{J}} \right]^2 + n + 2 }$.  
In the regime leading to the effective Hamiltonian in Eq.(\ref{sp121}), 
the large value of $U / t$ does not lead to
a Mott insulating phase, as it happens for a generic value of $f$. Indeed, 
the degeneracy between the states $ | n \rangle$ and $ |  n + 1 
\rangle$ at each site allows for restoring superfluidity, similarly to what
happens in the phase model describing one-dimensional arrays of Josephson
junctions  at the charge-degenerate point \cite{bradley_84}. 
 
Notice that the spin-1/2 Hamiltonian in Eq.(\ref{sp121}) has to be supplemented 
with the condition that $\sum_j S_j^z=0$, implying that 
physically acceptable states are only the eigenstates of   $\sum_j n_j$
belonging to the eigenvalue  $N_T$: this corresponds to singling out of 
the Hilbert space only the  zero magnetization sector. 
As discussed in detail in Ref.[\onlinecite{grst_13}], $H_{\rm spin-1/2} $ 
provides an excellent effective description of the low-energy dynamics of 
the BH model at half-odd filling. Although the mapping 
is done in the large-$U$ limit, in  Ref.[\onlinecite{grst_13}] it is shown that 
it is in remarkable agreement with DMRG results also for $U/J$ as low 
as $\sim 3-5$ and for low values of $N_T$ such as $N_T \sim 30$. 

Additional on-site energies $\epsilon_i$ can be accounted for by adding  
a term $\sum_{ j = - \ell}^{\ell} \epsilon_j n_j$ to the right-hand side of Eq.(\ref{bhh1}).
Accordingly, $H_{\rm spin-1/2}$ in Eq.(\ref{sp121}) has to be modified by adding 
the term $\sum_{ j = - \ell}^{\ell} \epsilon_j S_j^z$. As soon as the 
potential energy scale is smaller than $U$, we expect  
the mapping to be still  valid (we recall  that with a trapping parabolic 
potential typically  $\epsilon_j=\Omega j^2$ with 
$\Omega \equiv  m \omega^2 \lambda^2/8$, 
$m$ being the atom mass, $\omega$ the confining frequency 
and $\lambda/2$ the lattice spacing \cite{cataliotti01}). Yet, we 
stress that recent progresses in the realizations 
of potentials with hard walls \cite{gaunt12,mukherjee16} make the 
optical lattice realization of chains with open boundary conditions 
to lie within the reach of present technology.  

Another point to be addressed is what happens slightly away  from half-filling, that is,
for  $f=n+1/2+\varepsilon$, with $\varepsilon \ll 1$. In this case,   
one again recovers the effective Hamiltonian in Eq.(\ref{sp121}), but now with the 
constraint on physically acceptable states given by  $(1/{\cal N}) \langle \sum_j S_j^z \rangle=\varepsilon$. 
Since  keeping within  a finite magnetization sector is equivalent to having a nonzero applied magnetic field 
\cite{korepin_book}, one has then to add to the right hand side of 
Eq.(\ref{sp121}) a term of the form ${\cal H} \sum_j S_j^z$, where ${\cal H} \propto \varepsilon$: again, we expect that 
the mapping is valid as soon as that the magnetic energy is smaller 
than the interaction energy scale $U$, and, of course, that the system spectrum  
remains gapless \cite{affleck91}.

\begin{figure}[h]
\centering
\includegraphics*[width=.85\linewidth]{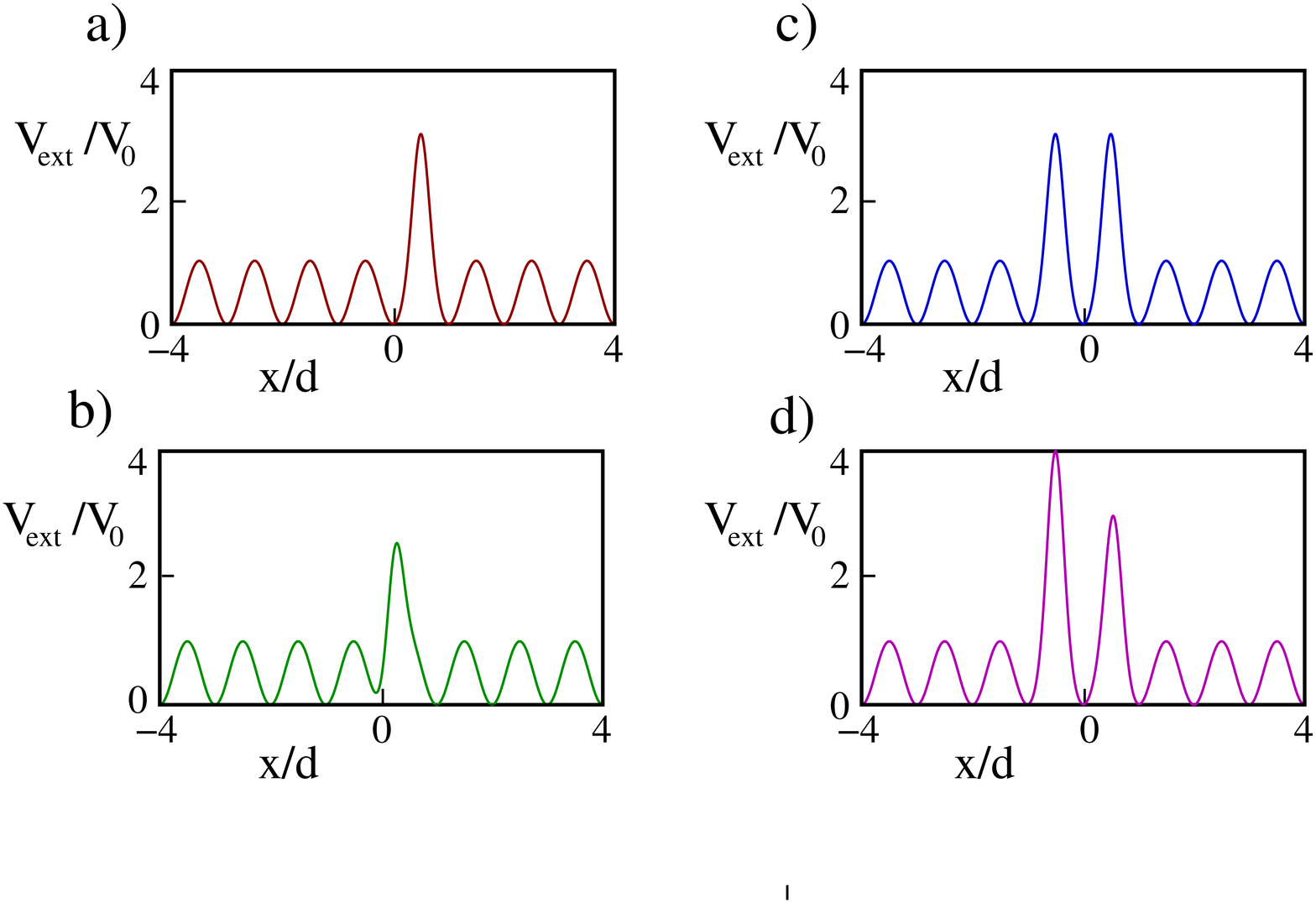}
\caption{External potential $V_{ext}$ (in units of $V_0$) as a function 
of $x$ (in units of $d$) for different values 
of $V_1/V_0$ and $V_2/V_0$ (with $\sigma=0.2 d$). 
In panels {\bf a)} and {\bf b)},  we consider $V_2=0$ so that only two hopping parameters are 
altered: panel {\bf a)}  corresponds to $x_1/d=0.5$  and 
panel {\bf b)} to $x_{0;1}/d=0.25$ (in both cases $V_1/V_0=2)$. In panels {\bf c)} and {\bf d)},  
we have $x_{0;1}/d=0.5$ and $x_{-1;0}/d=-0.5$ in both cases, but 
$V_1/V_0=V_2/V_0=2$ for panel {\bf c)} and $V_1/V_0=2$, $V_2/V_0=3$ 
for panel {\bf d)}.}
\label{pot}
\end{figure}

To modify the Hamiltonian in Eq.(\ref{sp121}) by adding bond impurities to the effective spin chain, 
we now create a link defect in the BH Hamiltonian in Eq.(\ref{bhh1}) by making  use of the fact that 
optical lattices provide a highly controllable 
setup in which it is possible to vary the parameters of the Hamiltonian as well 
as to add impurities with tunable parameters \cite{morsch06,bloch08}. This allows for 
creating a link defect in an optical lattice by  either   pertinently 
modulating the lattice, so that the energy barriers among 
its wells vary inhomogeneously across the chain, or by 
inserting one, or more, extra laser beams, centered on the minima of the lattice
potential. In this latter case, one makes the atoms feel a total potential 
given by $V_{ext}=V_{opt}+V_{laser}$, where 
the optical potential is given by $V_{opt}=V_0 \sin^2{(kx)}$, with 
$k=2 \pi / \lambda$ and 
$\lambda=\lambda_0 / \sin{\left( \theta/2 \right) }$, 
$\lambda_0$ being the wavelength of the lasers and $\theta$ the angle
between the laser beams forming the main lattice 
\cite{morsch06} (notice that the lattice spacing is $d=\lambda/2$). 
For counterpropagating laser beams having the same direction, $\theta=\pi$ 
and $d=\lambda_0/2$, while $d$ can be enhanced by making the
beams intersect at an angle $\theta \neq \pi$.  
$V_{laser}$ is the additional potential due to extra (blue-detuned) lasers: 
with one additional laser, 
centered at or close to an energy maximum of $V_{opt}$, 
say at $x \equiv x_{0;1}$ among the 
minima $x_0=0$ and $x_1=d$, the potential takes the form 
$V_{laser} \approx V_1 e^{-(x-x_{0;1})^2/\sigma^2}$. 
When the width $\sigma$ is much smaller 
than the lattice spacing, the hopping rate between the sites
$j=0$ and $j=1$ is reduced and no on-site energy term appears, 
as shown in panels {\bf a)} and {\bf b)} of Fig.\ref{pot}. Notice that we 
use a notation such that the $j$-th minimum corresponds 
to the minimum $x_j=jd$ in the continuum space. 

When $x_{0;1}$ is equidistant from  the lattice minima $x_0$ and $x_1$, 
corresponding to $x_{0;1}=\lambda/4=d/2$ and $\sigma < d$, 
then only the hopping $t_{0,1}$ is practically altered 
(see Fig.\ref{pot}{\bf a)} ). When $x_{0;1}$ 
is displaced from $d/2$ one has an asymmetry and also a nearest 
neighboring link [e.g., $t_{-1;0}$  in Fig.\ref{pot} {\bf b)}] may be altered 
(an additional on-site energy $\epsilon_0$ is also present). With $d \sim 2-3 \mu m$, 
one should have $\sigma \lesssim 2 \mu m$, in order to basically alter  
only one link. Notice that barrier of few $\mu m$ can be rather 
straightforwardly implemented \cite{albiez05,buccheri16} and 
recently a barrier of $\sim 2\mu m$ has been realized in  a Fermi gas 
\cite{valtolina15}. As discussed in the following, 
this is the prototypical realization of a weak-link impurity in an
otherwise homogeneous spin chain \cite{glazman97,giuliano05}. 

In general, reducing the hopping rate between links close to each other 
may either lead to an effective weak link impurity, or 
to a spin-1/2 effective magnetic
impurity, depending on whether the number of lattice sites 
between the reduced-hopping-amplitude 
links is even, or odd (see appendix \ref{kondo_hamiltonians} 
for a detailed discussion of this point). To ''double'' the construction
displayed in panels {\bf a)} and {\bf b)} of  Fig.\ref{pot}   to the one we 
sketch in panels {\bf c)} and {\bf d)} of Fig.\ref{pot}, we consider  a potential 
of the form $V_{laser} \approx V_1 e^{-(x-x_{0;1})^2/\sigma^2}+V_2 
e^{-(x-x_{-1;0})^2/\sigma^2}$ with $x_{-1;0}$ lying between 
sites $j=-1$ and $j=0$: assuming again 
$\sigma \lesssim d$, 
when $V_1=V_2$ and $x_{0;1}=-x_{-1;0}=d/2$ 
then only two links are altered, and in an equal way 
[the hoppings $t_{-1;0}$ and $t_{0;1}$ in  
Fig.\ref{pot} {\bf c)}], otherwise one has two different 
hoppings [again $t_{-1;0}$ and $t_{0;1}$ in   Fig.\ref{pot} {\bf d)} ].
When  $\sigma$ is comparable with $d$, apart from the variation of the 
hopping rates,  on-site energy terms enter the 
Hamiltonian in Eq.(\ref{bhh1}), giving rise to local magnetic fields in the spin 
Hamiltonian  in Eq.(\ref{sp121}). Though this latter kind of 
``site defects'' might readily be accounted for within the spin-1/2 $XXZ$ framework,
for simplicity we will not consider them in the following, and will only 
retain link defects, due to inhomogeneities in the boson hopping
amplitudes between nearest neighboring sites and in the interaction
energy $V$. Correspondingly, the 
hopping amplitude $t_{j;j+1}$ in Eq.(\ref{bhh1}) takes a dependence on the site  $j$
 also far form the region in which the potential $V_{laser}$ is 
centered.

In the following, we consider inhomogeneous distributions of link
parameters symmetric about the center of the chain (that is, about 
$j=0$). Moreover, for the sake of simplicity, we discuss 
a situation in which two (symmetrically placed) inhomogeneities
enclose a central region, whose link parameters may, or may not, be equal to the
ones of the rest of the chain. We believe that, though 
experimentally  challenging, this setup would correspond to the a situation 
in which the experimental detection of the Kondo length is cleaner.
In fact, we note that  all the experimental required  
ingredients are already available, as our setup requires two lasers 
with $\sigma \lesssim d$ (ideally,  $\sigma \ll d$) 
and centered  with similar precision.

As we discuss in detail in Appendix \ref{kondo_hamiltonians}, 
an ``extended central region'' as such can either be mapped onto
an effective weak link, between two otherwise homogeneous ``half-chains'', 
or onto an  effective isolated spin-1/2 impurity, weakly connected to the two 
half-chains. In particular, in this latter case, the Kondo effect may arise,
yielding remarkable nonperturbative effects and, eventually, ``sewing 
together'' the two half chains, 
even for a repulsive bulk interaction \cite{eggert92,furusaki98}. 
Denoting by ${\bf G}$ the region singled out
by weakening one or more links, 
in order to build an effective description of {\bf G}, we assume
that the mapping onto a spin-1/2 $XXZ$-chain works equally well with the
central region, and employ a systematic Shrieffer-Wolff (SW) summation, in order to trade
the actual dynamics of {\bf G} for an effective boundary Hamiltonian, that
describes the effective degrees of freedom of the central region interacting
with the half chains. One is then led to consider the Hamiltonian 
in Eq.(\ref{bhh1}), with link-dependent hopping rates $t_{ j ; j + 1}$. 

To illustrate how the mapping works, we focus onto the case of ${\cal M}=2$  altered links, 
corresponding  to two blue-detuned lasers, 
and briefly comment on the more general case. To resort to the  Kondo-like 
Hamiltonian for a spin-1/2 impurity embedded within a spin-1/2 $XXZ$-chain, 
we define the hopping rate to be equal to $t$ throughout the whole chain but 
between $j=-1$ and $j=0$, where we assume it to be equal to  $t_L$, and 
between $j=0$ and $j=1$, where we set it equal to $t_R$, 
corresponding to panels {\bf c)} and {\bf d)} of Fig.\ref{pot}.
On going through the SW transformation, one therefore
gets the  effective spin-$1/2$ Hamiltonian $H_{s}=H_{bulk}+H_K$, with  
$H_{bulk}=H_L+H_R$ and 
\begin{eqnarray}
 H_L &=& - J \sum_{ j = - \ell}^{-2} \left( S_j^+ S_{j+1}^- 
+ S_{j+1}^+ S_j^- \right) + 
J \Delta  \sum_{ j = - \ell}^{-2} S_j^z S_{j+1}^z \nonumber \\
 H_R &=& - J \sum_{ j = 1}^{\ell - 1} \left( S_j^+ S_{j+1}^- 
+ S_{j+1}^+ S_j^- \right) + 
J \Delta  \sum_{ j = 1}^{\ell - 1 } S_j^z S_{j+1}^z 
\;\;\;\; .
\label{lr_ham}
\end{eqnarray}
\noindent
The ''Kondo-like'' term is instead given by  
\begin{equation}
H_{K}=-J^{'}_{L} 
(S^{+}_{-1} S^-_{0} + S^{-}_{-1} S^{+}_{0} ) - 
J^{'}_{R} 
(S^{+}_{0} S^-_{1} + S^{-}_{0} S^{+}_{1} )
+J^{'}_{zL} S^z_{-1} S^z_{0} +J^{'}_{zR} S^z_{0} S^z_{1},
\label{HAM-K}
\end{equation}
where $J^{'}_{\alpha}=t_\alpha f$ and $J^{'}_{z \alpha}\approx V-3 
J^{'2}_{\alpha}/4U$ (with $\alpha=L,R$). 

Our choice for $H_K$ corresponds to 
the simplest case in which {\bf G} contains an even number 
of links -- or, which is the same, an odd number of sites, as 
schematically depicted 
in  Fig.\ref{impspin}{\bf b)}. We see that the isolated site works as 
an isolated spin-1/2 impurity ${\bf S}_{\bf G}$, 
interacting with the two half chains via
the boundary interaction Hamiltonian $H_{\bf B}^{(1)} \equiv H_K$.  
The other possibility, which we show in Fig.\ref{impspin}{\bf a)}, 
corresponds to 
the case in which  an odd number of links is altered and ${\bf G}$ 
contains an even number of sites. In particular, 
in Fig.\ref{impspin}{\bf a)} 
we have only one altered hopping coefficient. This latter case  is basically equivalent to a simple weak link between the 
$R$- and the $L$- half chain, which is expected to realize the spin-chain
version of Kane-Fisher physics of impurities in an interacting 
one-dimensional electronic system \cite{kane92}. 
In Appendix \ref{kondo_hamiltonians},
we review the effective low-energy description for a region {\bf G} containing
an in principle arbitrary number of sites. In particular, we conclude that  either 
the number of sites within {\bf G} is odd, and 
therefore the resulting boundary Hamiltonian takes 
the form of $H_K$ in Eq.(\ref{HAM-K}), or it is even, 
eventually leading to a weak link
Hamiltonian \cite{glazman97,giuliano05}. Even though this latter case is certainly 
an interesting subject of investigation, we are mostly interested 
in the realization of effective magnetic impurities. Therefore, henceforth we will be using  
$H_s$ as the main reference Hamiltonian, 
to discuss the emergence of Kondo physics in our system.

\section{Renormalization group flow of the impurity Hamiltonian parameters}
\label{ren_group}

\begin{figure}
\includegraphics*[width=.65\linewidth]{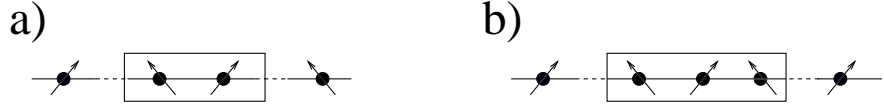}
\caption{Sketch of two different kinds of central regions in an otherwise 
uniform spin chain, respectively
realizing an effective weak-link impurity ({\bf a)}), and an effective 
spin-1/2 impurity ({\bf b)}).}
\label{impspin}
\end{figure}

In this section, we employ the renormalization group (RG) approach to recover the low-energy long-wavelength 
physics of a Kondo impurity in an otherwise homogeneous chain. From the RG equations we 
derive the formula for the invariant length which we eventually identify with $\xi_K$. In general, there are two standard ways 
of realizing the impurity in a spin chain, which we sketch in  Fig.\ref{impspin}. Specifically, 
we see that the impurity can be realized as an island    containing either an 
even or odd number of spins. The former case   is  equivalent to a weak link 
in an otherwise homogeneous chain, originally discussed in 
Refs.[\onlinecite{kanelett,kane92}]
for electronic systems, and  reviewed in detail in Ref.[\onlinecite{eggert99}] in
the specific context of spin chains. In this 
case, which we briefly review in Appendix \ref{cor_functions}, when $\Delta > 0$ in Eqs.(\ref{lr_ham}),
the impurity  corresponds to an irrelevant perturbation, which implies an 
RG flow of the system towards the 
fixed point corresponding to two disconnected chains, while for $\Delta < 0$
the weak link Hamiltonian becomes a relevant perturbation. 
Though this implies the emergence of an ''healing length '' 
for the weak link as an RG invariant length scale, 
with a corresponding flow towards a fixed point corresponding to 
the two chains joined into an effectively homogeneous single chain, 
there is no screening of a dynamical spinful impurity 
by the surrounding spin degrees of freedom and, accordingly, 
no screening cloud is detected in this case \cite{eggert99}. 

At variance,   a dynamical effective impurity screening takes place in the case of an 
effective spin-1/2  impurity \cite{affleck_length}. In this latter case, at any $\Delta$ 
such that $-1< \Delta < 1$,  the perturbative RG  approach shows that 
the disconnected-chain weakly coupled fixed point is ultimately unstable. 
In fact, the RG  trajectories flow towards 
a strongly coupled fixed point, which  we identify with the spin chain two channel Kondo fixed point, 
corresponding to healing  the chain but, at variance with what happens at a weak link 
for $0< \Delta \leq 1$, this time with  the chain healing taking place through an effective Kondo-screening 
of the magnetic impurity  \cite{eggert92}.  
 
A region containing an odd number of sites typically 
has a twofold degenerate
groundstate and, therefore, is mapped onto an effective 
spin-1/2 impurity ${\bf S}_{\bf G}$. 
The corresponding impurity Hamiltonian in Eq.(\ref{eo12}) takes the form of 
the Kondo spin-chain interaction Hamiltonian for a central 
impurity in an otherwise uniform spin chain 
\cite{sorensen}. To employ the bosonization formalism of appendix \ref{cor_functions} to 
recover the RG flow of  the impurity 
coupling strength, we resort to Eq.(\ref{bbi3}), corresponding to 
the bosonized spin Kondo Hamiltonian $H_K$ given by 
\begin{equation}
H_K =  \sum_{ \alpha =L,R} \left\{ - J_{\alpha}^{'} [ S_0^+ e^{ - 
\frac{i}{\sqrt{2}} \Phi_\alpha ( 0 ) } + S_0^- e^{  
\frac{i}{\sqrt{2}} \Phi_\alpha ( 0 ) } ] + J_{z\alpha}^{'} S_0^z \frac{ 
1}{\sqrt{2} \pi}
\frac{ \partial \Theta_\alpha ( 0 )}{ \partial x } \right\}
\:\:\:\: . 
\label{ksc30}
\end{equation}
\noindent
The  RG equations describing the flow  of the impurity coupling strength can be
derived by means of standard 
techniques for  Kondo effect in spin chains \cite{furusaki98} and, in particular, by 
considering the fusion rules between the various operators entering 
$H_K$ in Eq.(\ref{ksc30}). In doing so, in principle additional, 
weak link-like, operators 
describing direct tunneling between the two chains can be generated, 
such as, for 
instance, a term $\propto 
e^{ \frac{i}{\sqrt{2}} [ \Phi_L ( 0 ) - \Phi_R ( 0 ) ]}$, with scaling dimension 
$h_A = \frac{1}{g}$. However, one may safely neglect 
a term as such, since, for $g<1$, 
it corresponds to an additional irrelevant boundary 
operator that has no effects on the RG flow of the 
running couplings appearing in $H_K$. 
For $g \geq 1$ it becomes marginal, or relevant, but still subleading, 
compared to the terms $\propto J_\alpha^{'}$, 
as we discuss in the following and, therefore, it can 
again be neglected for the purpose of working out the RG 
flow of the boundary couplings. This observation effectively 
enables us to neglect operators mixing the 
$_L$ and the $_R$ couplings with each other and, accordingly, 
to factorize the RG equations for the running couplings with respect to the index 
$_\alpha$. 

More in detail,  we define the dimensionless variables
$G_\alpha ( \ell )$ and $G_ {z , \alpha  } ( \ell )$ as 
\begin{equation}
G_{\alpha} (\ell) = \left( \frac{\ell}{\ell_0} \right)^{1 - \frac{1}{2g}} 
\frac{J_{\alpha}^{'}}{J} \, \, \, \, \, \, \, \, 
{\rm and} \, \, \, \, \, \, \, \, G_{z , \alpha} ( \ell ) = 
\frac{J_{z\alpha}^{'}}{J}
\:\:\:\: , 
\label{dimensionless}
\end{equation} 
\noindent
(see Appendix \ref{boso_imp} for a discussion on the estimate of the reference length $\ell_0$) 
with $\alpha = L , R$. 

The RG  equations for the running couplings are 
given by
\begin{eqnarray}
\frac{ d G_{\alpha} ( \ell  ) }{ d \ln ( \frac{\ell}{\ell_0} )} &=& 
h_g G_{\alpha }  ( \ell ) + G_{\alpha}  (\ell )  G_{z , \alpha}  (\ell )
\nonumber \\
\frac{ d G_{z \alpha} ( \ell ) }{ d \ln ( \frac{\ell}{\ell_0} )} &=&
G_{\alpha}^2 ( \ell )
\:\:\:\: ,
\label{ksc31}
\end{eqnarray}
\noindent 
with $h_g = 1 - 1 / (2g)$. 
For the reasons discussed above, the RG  equations 
in Eq.(\ref{ksc31}) for the $_L$- and the $_R$-coupling 
strengths are decoupled from each other. In fact, 
they are formally identical to  the corresponding equations obtained 
for a single link impurity placed at the end of the chain 
(``Kondo side impurity'') \cite{sorensen}. At variance with this 
latter case, as argued by Affleck and Eggert \cite{eggert92}, in our specific case of 
a ''Kondo central impurity'' the scenario for what 
concerns the possible Kondo-like fixed points is much richer, according 
to whether $G_{L} (\ell_0 ) 
\neq G_{R} ( \ell_0 )$ (``asymmetric case''), or  $G_{L} (\ell_0 ) 
= G_{R} ( \ell_0 )$ (``symmetric case''), as we discuss
below. 

To integrate Eqs.(\ref{ksc31}),  we define the reduced variables 
$X_\alpha (\ell) \equiv G_\alpha(\ell)$ and $X_{z , \alpha} ( \ell ) = 
G_{z, \alpha} ( \ell ) + 1 - \frac{1}{2 g}$ 
for $\alpha=L,R$ (since the equations for the two values of $\alpha$ are 
 formally equal to each other, from now on
we will understand the index $\alpha$).
As a result, one gets
\begin{equation}
  \frac{ d X ( \ell ) }{ d \ln ( \frac{\ell}{\ell_0} )} =
X (\ell ) X_{z} ( \ell );\, \, \, \, \, \, \, \,
\frac{ d X_{z} ( \ell ) }{ d \ln ( \frac{\ell}{\ell_0} )} =
X^2 ( \ell ). 
\label{ksc15}
\end{equation}
\noindent
Equations  (\ref{ksc15}) coincide with the RG equations 
obtained for the Kosterlitz-Thouless phase transition \cite{itzykson89}.
To solve them, we note that the quantity
\begin{equation}
\kappa = X_z^2 ( \ell ) - X^2 (\ell )  
\;\;\;\; , 
\label{ksc16}
\end{equation}
\noindent
is invariant along the RG trajectories. In terms of the microscopic parameters
of the BH Hamiltonan one gets $\kappa=\kappa(\ell_0)=
(V/J-3J^{'2}/(4UJ) +1-1/(2g))^2 - (J^{'}/J)^2$. 
To avoid the onset of Mott-insulating phases, we have 
to assume that the interaction is such that $g > 1/2$. This implies  
$h_g > 0$ and $X_z (\ell_0) >0$: thus, we assume 
$X (\ell_0 ), X_z ( \ell_0 )>0$. 
This means that the RG trajectories  always lie 
within the first quarter of the $(X,X_z)$-parameter plane and, in particular, that the 
running couplings always grow along the trajectories. 

\begin{figure}
\includegraphics*[width=.8\linewidth]{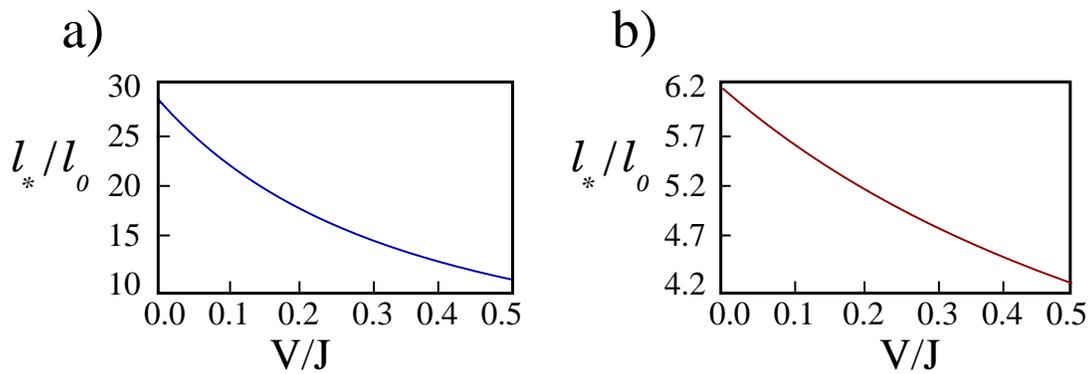}
\caption{$\ell_* / \ell_0$ as a function of $V/J$ for 
$0 \leq V/J \leq 0.5$. The 
other parameters are chosen so that $U/J=4$ and 
$J'/ J = 0.2$ (panel {\bf a)}), and 
$J' / J = 0.6$ (panel {\bf b)}). As discussed in Appendix \ref{boso_imp}, $\ell_0$ is of order 
of the lattice spacing $d$.} \label{jpr}
\end{figure}

Using the constant of motion in Eq.(\ref{ksc16}), Eqs.(\ref{ksc15}) can be
easily integrated. As a result, one may estimate the RG invariant length scale $\ell_*$ 
defined by the condition that, at the scale $\ell \sim \ell_*$, the perturbative
calculation breaks down (which leads us to eventually identify $\ell_*$ with 
$\xi_K$). As this is signaled by the onset of a divergence in 
the running parameter $X ( \ell )$ \cite{giu_stt}, one may find the explicit 
formulas for $\ell_*$, depending on the sign of $\kappa$, as detailed below:

\begin{itemize}
 \item {\bf $\kappa=0$}. In this case, as the symmetry at $\ell = \ell_0$ 
between $K $ and $X_z$ is preserved along the RG 
trajectories, it is enough to provide the explicit solution for 
$X_z ( \ell ) (= X ( \ell ) )$, which is 
given by 
\beq
X_z (\ell) = \frac{X_z (\ell_0 )}{ 1 - X_z ( \ell_0 ) \ln ( \frac{\ell}{\ell_0} ) }
\:\:\:\: . 
\label{rgflow.1}
\eneq
\noindent
From Eq.(\ref{rgflow.1}), one obtains 
\begin{equation}
\ell_* \sim \ell_0 \exp \left[  \frac{1}{X_z (\ell_0 ) } \right]
\:\:\: ,
\label{lstar.k0}
\end{equation}
\noindent
which is the familiar result one recovers for the ''standard'' Kondo
effect in metals \cite{affleck_length}.

\item {\bf  $\kappa<0$}. In this case, the explicit solution of 
Eqs.(\ref{ksc15}) is given by 
\begin{eqnarray}
 X_z ( \ell ) &=& \sqrt{- \kappa } \tan \left\{ {\rm atan} \left[ \frac{X_z ( \ell_0 ) }{\sqrt{- \kappa}} \right] + 
 \sqrt{- \kappa} \ln \left( \frac{\ell}{\ell_0} \right) \right\} \nonumber \\
 X ( \ell ) &=& \sqrt{-\kappa + X_z^2 ( \ell ) }
 \:\:\:\: , 
 \label{lstar.k1}
\end{eqnarray}
\noindent
which yields 
\beq
\ell_* \sim \ell_0 \exp \left[ \frac{ \pi  - 2 \:
 {\rm atan} ( \frac{X_z (\ell_0 ) }{\sqrt{ | \kappa | } } ) }
{2 \sqrt{ | \kappa | }} \right]
\:\:\:\: . 
\label{lstar.k11}
\eneq
\noindent

\item {\bf $\kappa > 0$}. In this case one obtains 
\begin{eqnarray}
 X_z ( \ell ) &=& - \sqrt{\kappa} \left\{ \frac{[ X_z ( \ell_0 )  - \sqrt{\kappa}] \left( \frac{\ell}{\ell_0} \right)^{2\sqrt{\kappa}} + [ X_z ( \ell_0 ) + \sqrt{\kappa} ]  }{
 [ X_z ( \ell_0 )  - \sqrt{\kappa} ] \left( \frac{\ell}{\ell_0} \right)^{2\sqrt{\kappa}} - [ X_z ( \ell_0 ) + \sqrt{\kappa} ] } \right\} \nonumber \\
 X ( \ell ) &=& \sqrt{-\kappa + X_z^2 ( \ell ) }
 \:\:\:\: .
 \label{lstar.k12}
\end{eqnarray}
\noindent
As a result, we obtain 
\beq
\ell_* \sim \ell_0 \left\{    \frac{X_z (\ell_0 ) + \sqrt{\kappa}}{
X_z (\ell_0 ) - \sqrt{\kappa} }  \right\}^\frac{1}{2 \sqrt{\kappa}}
\:\:\:\: . 
\label{lstar.k13}
\eneq
\noindent
\end{itemize}
To provide some realistic estimates of $\ell_*$, 
in Fig.\ref{jpr} we plot $\ell_* / \ell_0$ as 
a function of the repulsive interaction potential $V$, keeping fixed 
all the other 
system parameters (see the caption for the numerical values 
of the various parameters). 
The two plots we show correspond to different values of $J^{'}$. 
We see that, as expected, at any value of $V / J$, $\ell_*$ 
decreases on increasing $J^{'}$. We observe that with realistically 
small values of $V/J$, say between $0$ and $0.5$, one has a value of the Kondo 
length order of $20$ sites (for $J'/J=0.2$) and $5$ sites (for $J'/J=0.6$), 
that should detectable from experimental data.

Also, we note a remarkable decrease of $\ell_*$ with $ V / J$ and, 
in particular, a finite 
$\ell_*$ even at extremely small values of $V$, which correspond to 
negative values of $J_z^{'}$ and, thus, 
to an apparently ferromagnetic Kondo coupling between the impurity 
and the chain. In fact, in order for the Kondo coupling to be 
antiferromagnetic, and, thus, to correspond to a relevant boundary perturbation, 
one has to either have both 
$J^{'}$ and $J_z^{'}$ positive, or 
the former one positive, the latter negative.
In our case, the RG equations in 
Eqs.(\ref{ksc15}),  show how the $\beta$-function for the running 
coupling $X (= G) $ is proportional to $X_z G$, rather than to 
$G_z G$. Thus, what matters here 
is the fact that $X_z - G_z = 1 - \frac{1}{2g} > 0$,
which makes $X_z ( \ell_0 ) $ positive even though $G_z ( \ell_0 )$ 
is negative. As a result, even when both $J^{'}$ and $J_z^{'}$ are 
negative as it may happen, for 
instance, if one starts from a BH model with $V \sim 0$, one may 
still recover a Kondo-like RG flow and find a 
finite $\ell_*$, as evidenced
by the plots in Fig.\ref{jpr}.

\noindent
Being an invariant quantity along the RG trajectories, here $\ell_*$  plays the same role as   $\xi_K$ in  the 
ordinary Kondo effect, that is, once the RG  trajectories 
for the running strengths are constructed by using the system size $\ell$ as driving variable, 
all the curves are expected to 
collapse onto each other, provided that, at each curve, $\ell$ is rescaled 
by the corresponding $\ell_*$
\cite{affleck_length,hewson,sholl,florens15}. 

In fact, in the specific type of system we are 
focusing onto, that is, 
an ensemble of cold atoms loaded on a pertinently 
engineered optical lattice, it may be difficult to
vary $\ell$ by, in addition, 
keeping the filling constant (not to affect the 
parameters of the effective Luttinger liquid model Hamiltonian 
describing the system). Yet, one may resort to a fully complementary 
approach in which, as we highlight in the following, the length $\ell$, as well as the filling $f$, 
are kept fixed and, taking advantage of the scaling 
properties of the Kondo RG flow, one probes the 
scaling properties by 
varying $\ell_*$. Indeed, from our 
Eqs.(\ref{lstar.k0},\ref{lstar.k11},\ref{lstar.k13}), one sees 
that in all cases of interest, the relation between $\ell_*$ and 
the microscopic parameters 
characterizing the impurity Hamiltonian is known. As a result, 
one can in principle arbitrarily
tune $\ell_*$ at fixed $\ell$ by varying the tunable system 
parameter. As we show in the following, 
this provides an alterative way 
for probing scaling behavior, more suitable to an optical lattice 
hosting a cold atom condensate. 
In order to express the integrated RG flow equations 
for the running parameters
as a function of $\ell$ and $\ell_*$, it is sufficient to integrate the 
differential equations 
in Eqs.(\ref{ksc15}) from $\ell_*$ up to $\ell$. As 
a result, one obtains the following equations:
\begin{itemize}
 \item {\bf For $\kappa = 0$}: 
 \beq
X ( \ell ) = X_z (\ell) = \frac{X_z ( \ell_0) }{ -  \ln ( \frac{\ell}{\ell_*} ) }
\:\:\:\: ; 
\label{rgflowx.1}
\eneq
\noindent
\item {\bf For $\kappa < 0$}:
\begin{eqnarray}
 X_z ( \ell ) &=& \sqrt{- \kappa } \tan \left\{ \frac{\pi}{2} - 
 \sqrt{-\kappa}   \ln \left( \frac{\ell_*}{\ell} \right) \right\} \nonumber \\
 X ( \ell ) &=& \sqrt{-\kappa + X_z^2 ( \ell ) }
 \:\:\:\: ; 
 \label{lstarx.k1}
\end{eqnarray}
\noindent
\item {\bf For $\kappa > 0$}:
\begin{eqnarray}
 X_z ( \ell ) &=&   \sqrt{\kappa} \left\{ \frac{\left( \frac{\ell_*}{\ell} \right)^{2\sqrt{\kappa}} + 
  }{ \left( \frac{\ell_*}{\ell} \right)^{2\sqrt{\kappa}} -1 } \right\} \nonumber \\
 X ( \ell ) &=& \sqrt{-\kappa + X_z^2 ( \ell ) }
 \:\:\:\: . 
 \label{lstarx.k12}
\end{eqnarray}
\noindent
\end{itemize} 

From Eqs.(\ref{rgflowx.1},\ref{lstarx.k1},\ref{lstarx.k12}), one 
therefore concludes that, once 
expressed in terms of $\ell / \ell_*$, the integrated RG flow for the running coupling 
strengths only depends on the parameter $\kappa$. Curves corresponding to 
the same values of $\kappa$ just 
collapse onto each other, independently of the values of all the 
other parameters. 

We pause here for an important comment. As discussed in \cite{sorensen}, in the 
spin chain realization of the Kondo model,  one exactly retrieves 
the equation of the conventional Kondo effect  at $g=1/2$ only after
adding a frustrating second-neighbor interaction, thus resorting to 
the so-called $J_1- J_2$ model Hamiltonian. In principle, the same 
would happen for the $XXX$-spin chain with nearest-neighbor interaction only, 
except that, strictly speaking, the correspondence is exactly realized 
only in the limit of an infinitely long chains. In the case of finite chains, 
the presence of a marginally irrelevant Umklapp  operator may induce finite-size violations
from Kondo scaling which, as stated above, disappear in the thermodynamic limit. 
Yet, as this point is mostly of interest because it may affect the precision of 
numerical calculations, we do not address it here and refer to Ref.[\onlinecite{sorensen}]
for a detailed discussion of this specific topic.

Another important point to stress is that, strictly speaking, we have so far 
neglected the possible  effects of the asymmetry ($J_L^{'} \neq J_R^{'}$ and 
$J_{z , L}^{'} \neq J_{  z , R}^{'}$), versus symmetry ($J_L^{'} = J_R^{'}$ and 
$J_{z , L}^{'} = J_{  z , R}^{'}$) in the bare couplings. In fact, the   nature of the 
stable Kondo fixed point reached by the system in the large scale limit deeply depends on whether or not the bare
couplings between the impurity and the chains are symmetric, or not. 
Nevertheless, as we argue in the following, one sees that, while the nature of the Kondo fixed point may be quite 
different in the two cases (two-channel versus one-channel spin-Kondo 
fixed point), 
one can still expect to be able to detect the onset of the Kondo regime and 
to probe the corresponding Kondo  length 
by looking at the density-density 
correlations in real space, though the correlations themselves 
behave differently in the two cases. We discuss at 
length about this latter point in the next 
section. Here, we rather discuss  about 
the nature of the Kondo fixed point in 
the two different situations, starting with the case of symmetric 
couplings between the impurity and the chains. 

When $J_{L}^{'} = J_R^{'}$ and $J_{z  L}^{'} = J_{z  R}^{'}$, since, 
to leading order in the running couplings, 
there is no mixing between the $_L$- and the $_R$ coupling 
strengths, the $L-R$ symmetry is  not expected to be broken all the way down to the strongly coupled 
fixed point which, 
consequently, we identify with the two-channel spin-chain Kondo fixed 
point, in which the 
impurity is healed and the two chains have effectively joined into a 
single uniform chain.
Due to the $L-R$ symmetry, one can readily show that all the allowed 
boundary operators at the strongly coupled fixed point 
are irrelevant \cite{eggert92,furusaki98}, leading 
to the conclusion that the two-channel spin-Kondo fixed point is stable, in 
this case. 

Concerning the effects of the asymmetry, on  comparing the scale dimensions
of the various impurity boundary operators,   one expects them to be particularly relevant if the 
asymmetry is realized in the transverse Kondo coupling strengths, 
that is, if one has  $J_{L}^{'} \gg J_R^{'}$. We assume that this is the case which, moving to the 
dimensionless couplings, implies  
$G_L ( \ell_0 ) \gg G_R ( \ell_0 )$. Due to the monotonicity of 
the integrated RG curves, we expect that this 
inequality keeps preserved along the integrated flow, that is, 
$G_L ( \ell ) \gg G_R ( \ell )$ at any scale 
$\ell \geq \ell_0$. In analogy with  
the standard procedure used with multichannel Kondo effect with non-equivalent
channels, one defines 
$\ell_*$ as the scale at which the larger running coupling $G_L ( \ell )$ 
diverges, which is the signal of 
the onset of the nonperturbative regime. Due to the coupling asymmetry, we 
then expect $G_R ( \ell_* ) \ll 1$, that is, 
at the scale $\ell \sim \ell_*$, 
the system may be regarded as a semi-infinite chain at the left-hand side, 
undergoing Kondo effect with an isolated magnetic impurity, 
weakly interacting with a second semi-infinite chain, 
at the right-hand side. To infer the effects of 
the residual coupling, one may assume that, at 
$\ell \sim \ell_*$, the impurity is ``re-absorbed'' in the left-hand chain 
\cite{eggert92,furusaki98}, so that this scenario will consist of the left-hand chain, with one 
additional site, connected with a link of strength $\sim G_R ( \ell_* )$ 
to  the endpoint of the right-hand chain.  Within 
the bosonization approach, the weak link Hamiltonian is given by   \cite{kane92} 
\beq
V_B^{\rm Asym} \sim 
- G_R ( \ell_* )   e^{ \frac{i}{ \sqrt{2}} [ \Phi_L ( 0 ) - \Phi_R ( 0 ) ]} + {\rm h.c.}
\:\:\:\: .
\label{kf1}
\eneq
\noindent
$V_B^{\rm Asym}$ has scaling dimension $\frac{1}{g}$. Depending on whether  
$g > 1$, or $g<1$, it can therefore be either relevant, 
or irrelevant (or marginal
if $g=1$). When relevant, it drives the system towards a fixed point in 
which the 
weak link is healed. When irrelevant, the fixed point corresponds to the 
two disconnected chains.
In either case, the residual flow takes place after the onset of Kondo 
screening. We therefore conclude that Kondo screening takes 
place in the left-hand chain only 
and, accordingly, one expects to be able to probe $\ell_*$  by 
just looking at the real space density-density correlations in that chain only.
From the above discussion we therefore conclude that Kondo effect is actually 
realized at a chain with an effective spin-1/2 impurity whether or not 
the impurity couplings to the chains 
are symmetric, or not, though the fixed point the 
system is driven to along the RG 
trajectories can be different in 
the two cases. 
  
\section{Density-density correlations and measurement of the Kondo length}
\label{density_density}

In analogy to the screening length 
$\xi_K$ in the standard Kondo effect \cite{noz_1,noz_2},  in the spin chain realization of 
the   effect,    the   screening length 
$\ell_*$ is identified with the typical size of a cluster of spins 
fully screening the moment
of the isolated magnetic impurity, either lying at one side 
of the impurity itself 
(in the one-channel version of the effect-side impurity at the 
end of a single spin chain), 
or surrounding the impurity on both sides (two channel version
of the effect-impurity embedded within an otherwise uniform chain). 
  
So far, $\ell_*$ showed itself
as quite an elusive quantity to experimentally 
detect, both in electronic Kondo effect, 
as well as in spin Kondo effect \cite{Affleck09}.
In this section, we propose to probe 
$\ell_*$ in the effective 
spin-1/2 $XXZ$ chain describing the BH model, 
by measuring the integrated
real-space density-density correlation functions.  Real-space density-density correlations
in atomic condensates on an optical lattice can be measured with a good level 
of accuracy (see e.g. Refs.[\onlinecite{pethick02,ibloch_nature}].)  Given the mapping between the 
BH- and the spin-1/2 $XXZ$ spin Hamiltonian, real-space density-density correlation 
functions are related via Eq.(\ref{spinoperators}) to the  correlation functions of 
the $z$-component of the effective spin operators in the $XXZ$-Hamiltonian 
(local spin-spin susceptibility), which eventually enables us to analytically compute the correlation function within 
spin-1/2 $XXZ$ spin chain Hamiltonian framework. 
The idea of inferring informations on the Kondo length   
 by looking at the scaling properties of the real-space 
local spin susceptibility was  put forward in Ref.[\onlinecite{barzykin}].
In the specific context of lattice model Hamiltonians, 
the integrated real-space 
correlations have been proposed as a tool to extract $\xi_K$ in a 
quantum dot, regarded as a local Anderson model, interacting with 
itinerant lattice spinful fermions \cite{sholl}. 
Specifically, letting ${\bf S}_{\bf G}$ denote the spin
of the isolated spin-1/2 impurity and ${\bf S}_j$ the spin operator in 
the site $j$, assuming that the impurity is located 
at one of the endpoints of the 
chain and that the whole model, 
including the term describing the interaction between
${\bf S}_{\bf G}$ and the spins of the chain, is spin-rotational invariant, 
one may introduce the integrated real-space correlation function 
${\bf \Sigma} ( x )$, defined as 
\cite{sholl}
\beq
{\bf \Sigma} ( x ) = 1 + \sum_{ y = 1}^x \left[ \frac{\langle {\bf S}_{\bf G} \cdot {\bf S}_y \rangle }{\langle {\bf S}_{\bf G}
\cdot {\bf S}_{\bf G}\rangle } \right]
\:\:\:\: . 
\label{dd.1}
\eneq
\noindent
The basic idea is that the first zero of ${\bf \Sigma} ( x )$ 
one encounters in moving from the 
location of the impurity, identifies the portion of the whole chains 
containing the spins that  fully screen ${\bf S}_{\bf G}$. 
Once one has found the solution of the equation
${\bf \Sigma} ( x =  x_*  ) = 0 \: ,$ 
one therefore  naturally identifies $x_*$ with $\ell_*$.  
It is important to stress  that this idea equally applies 
whether one is considering the spin impurity at just one side of
the chain (one-channel 
spin chain Kondo), or embedded within the chain   (two-channel spin chain Kondo). 
Thus, while in the following we mostly consider the two-channel case,   we readily infer that our discussion 
applies also to the one-channel case.

\begin{figure}
\includegraphics*[width=1.\linewidth]{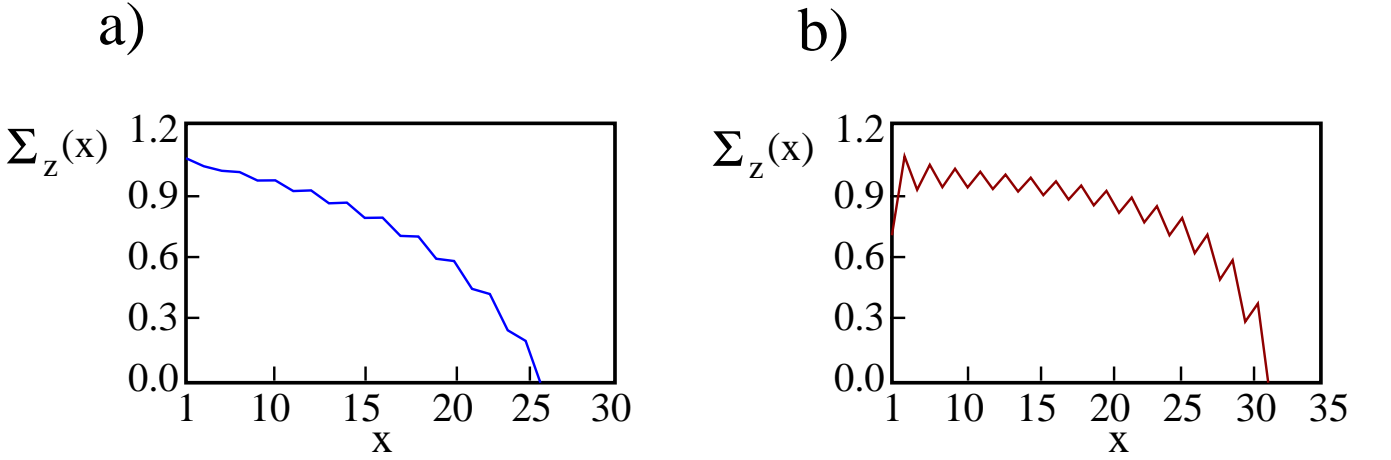}
\caption{{\bf a)}: Plot of ${\bf \Sigma}_z ( x )$ {\it vs.} 
$x$ for $U/J = 4$, $V=0$ 
(corresponding to $\Delta = - 0.1875$) and $J' / J = 0.2$. 
From the plot one infers 
$\ell_* \sim 26$, which is in good agreement with the value obtained from the plot in Fig.\ref{jpr}{\bf a)}. {\bf b)}: 
Same as before, but with $V / J =2.1875$ (corresponding to $\Delta = 0.2$) 
and $J' / J = 0.1$. As expected, the lower value of $J'$ yields a larger 
$\ell_* \sim 32$.} 
\label{correlationsc}
\end{figure}

To adapt the approach of  Ref.[\onlinecite{sholl}] to our specific case,
first of all, since our impurity is located at the center of 
the chain, one has to modify the definition of ${\bf \Sigma} ( x )$ 
so to sum over  $j$ running from $-x$ to $x$. 
In addition, in our case both the bulk spin-spin interaction, 
as well as the effective 
Kondo interaction with the impurity, are not isotropic in the spin space. 
This requires modifying the 
definition of ${\bf \Sigma} ( x )$, in analogy to what 
is done in Ref.[\onlinecite{sholl}] in the 
case in which an applied magnetic field 
breaks the spin rotational invariance. Thus, to probe 
$\ell_*$ we use   the integrated $z$-component of the spin correlation 
function, ${\bf \Sigma}_z ( x )$, defined as 
\beq
{\bf \Sigma}_z ( x ) = 1 + \sum_{ y = -x }^x \left[ \frac{\langle S_{\bf G}^z S_y^z \rangle - \langle S_{\bf G}^z \rangle \langle S_y^z \rangle}{
\langle S_{\bf G}^z S_{\bf G}^z \rangle -\langle S_{\bf G}^z \rangle^2  } \right]
\:\:\:\: . 
\label{dd.2}
\eneq
\noindent
In general, estimating $\ell_*$ from ${\bf \Sigma}_z ( x )$ 
would require exactly computing the spin-spin correlation functions 
by means of a numerical technique, such as it is done in 
Ref.[\onlinecite{sholl}] -- nevertheless one in general 
expects that the estimate of $\ell_*$ obtained using perturbative 
RG differs by a factor order of $1$ from the one 
obtained by nonperturbative, numerical means.  
For the purpose of  showing the consistency between the estimate of $\ell_*$ 
from the spin-spin correlation
functions and the results from the perturbative analysis of  
Sec.\ref{ren_group}, one therefore 
expects it to be sufficient to resort to a perturbative (in $J_z^{'} , J'$) 
calculation of 
${\bf \Sigma}_z ( x )$, eventually improved by substituting 
the bare coupling strengths with the running ones, 
computed at an appropriate scale \cite{affleck_length}. 
To leading order in the impurity couplings, we obtain 
\begin{eqnarray}
\langle S_{\bf G}^z S_y^z \rangle &=& -  J_{z , R}^{'}  \: \int_0^\infty \: d \tau \: 
G_{z,z} (y , 1 ; \tau | \ell )  \: , \; (y > 0 ) \nonumber \\
\langle S_{\bf G}^z S_y^z \rangle &=&  -  J_{z , L}^{'}  \: \int_0^\infty \: d \tau \: 
G_{z,z} (y , 1 ; \tau | \ell )  \: , \; (y < 0 )
\:\:\:\: , 
\label{dd.3}
\end{eqnarray}
\noindent
with the finite-$\tau$ correlation function 
$G_{z,z} ( x , x' ; \tau | \ell )$ defined 
in Eq.(\ref{corr.x2}). To incorporate 
scale effects in the result of Eq.(\ref{dd.3}), we
therefore replace the bare impurity coupling strengths with the running ones 
we derived in Sec.\ref{ren_group}, 
computed at an appropriate length scale, which we 
identify with the size $x$ of the spin cluster effectively 
contributing to impurity 
screening.  Therefore, referring to the 
dimensionless running coupling $X_z ( \lambda )$ defined in Eqs.(\ref{ksc31}), 
we obtain
\begin{eqnarray}
{\bf \Sigma}_z ( x ) &=& 1 - \frac{8 J_z^{'}  ( x ) \ell}{\pi u } \: \sum_{y=1}^x \:  \int_0^\infty \: d w \: G_{z,z} \left( y , 1 ;\frac{ \pi u w }{\ell} \biggr| \ell \right)   
\nonumber \\
&=& 1 - 8  \varphi ( \Delta ) \left[ X_z  ( x ) + \frac{1}{2g} - 1 \right] \: \ell   \: \sum_{y=1}^x \:  \int_0^\infty \: d w \: G_{z,z} \left( y , 1 ;\frac{ \pi u w }{\ell} \biggr| \ell \right)   
\:\:\:\: ,
\label{dd.5}
\end{eqnarray}
\noindent
with $\varphi ( \Delta )$ given by 
\beq
\varphi ( \Delta ) = \frac{ {\rm arcos} \left( \frac{\Delta}{2} \right) }{\pi^2  \sqrt{1 - \left( \frac{\Delta}{2} \right)^2}}
\:\:\:\: .
\label{dd.6}
\eneq
\noindent
Remarkably, $\varphi ( \Delta ) \to 1$ as $\Delta \to 0$. 
In Fig.\ref{correlationsc}, we 
show ${\bf \Sigma}_z ( x )$ {\it vs.} $x$ 
(only the positive part of the graph) for two 
paradigmatic situations: in Fig.\ref{correlationsc}{\bf a)} we consider 
the absence of nearest-neighbor ''bare'' density-density interaction ($V=0$). 
  In Fig.\ref{correlationsc}{\bf b)} we consider a rather large, presently 
not straightforward to be implemented in experiments, value 
of $V$ ($V/J \sim 2.2$) to show the results for the Kondo length 
with a positive value of the $XXZ$ anisotropy parameter. We see that 
there is not an important dependence of the Kondo length upon $V$, since 
the main parameter affecting $\ell_*$ is actually given by $J'/J$.

From the analysis of Ref.[\onlinecite{grst_13}], one sees that, 
even at $V=0$, a nonzero attractive density-density interaction 
between nearest-neighboring sites of the chain is actually induced by 
higher order (in $t /U$) virtual processes, 
which implies that, for $V=0$, $g$ keeps slightly higher than $1$. 
At variance, for finite $V$, $g$ can be either larger, 
or smaller than $1$, as it is the case in the plot 
in Fig.\ref{correlationsc}{\bf b)}. 
In both cases we see the effect of ''Friedel-like'' oscillations in the 
density-density correlation, 
which eventually conspire to set ${\bf \Sigma}_z ( x )$ 
to $0$ at a scale $x \sim \ell_*$ (see the caption of the figures
for more details on the numerical value of the various parameters). 

\noindent

In general, Eq.(\ref{dd.5}) has to be regarded within the context of the  
general scaling theory for ${\bf \Sigma}_z ( x )$ \cite{affleck_length}. 
In our specific case, at variance with what happens in 
the  ''standard'' Kondo problem of itinerant electrons in a metal 
magnetically interacting with an isolated impurity \cite{affleck_length}, 
the boundary action in Eq.(\ref{bbi3}) contains terms that are 
relevant as the length scale grows. In general, in this case 
a closed-form scaling formula for physical quantities cannot be inferred 
from the perturbative results, due to the proliferation of 
additional terms generated at 
higher orders in perturbation theory \cite{amit}. Nevertheless, here 
one can still recover a pertinently adapted scaling equation, 
as only dimensionless
contributions to $S^B_{\bf G}$ effectively contribute 
${\bf \Sigma}_z ( x )$ to any order in 
perturbation theory. The point is that, 
as we are considering a boundary operator in a bosonized theory in 
which the fields $\Phi _{L , R }( x , \tau )$ 
obey Neumann boundary conditions at the boundary, 
the fields $\Theta_{L . R } ( 0 , \tau )$ appearing in the bosonized formula for 
$S_{1 , L}^z , S_{1 , R }^z$ in Eqs.(\ref{core3bis}) 
are pinned at a constant for any $\tau$. As a 
result, the corresponding contribution to the boundary interaction 
reduces to the one in Eq.(\ref{bbi3}),
which is purely dimensionless and, therefore, marginal. 
As for what concerns the contribution
$\propto J^{'}_{L , R}$, it is traded for a marginal 
one once one uses as running couplings the rescaled variables 
$X_L$ and
$X_R$, rather than $J'_L , J'_R$. Now, from Eqs.(\ref{core3bis}) 
we see that the bosonization formula for $S_j^z$ contains a term 
that has dimension $d_1=1$ and a term with dimension $d_2 = (2g)^{-1}$. 
Taking into account the 
dynamics of the degrees of freedom of the chains comprised over a 
segment of length $x$, we therefore may 
make the scaling ansatz for ${\bf \Sigma}_z$ in the form
\beq
{\bf \Sigma}_z [ x , \ell , X_z , X ] = \: \tilde{\omega}_0  \left[ \frac{x}{\ell} , X_z , X \right] + 
\ell^{ 1-g} \: \tilde{\omega}_1 \left[ \frac{x}{\ell} , X_z , X \right]
\:\:\:\: , 
\label{dd.7}
\eneq
\noindent
with $\omega_0 , \omega_1$ scaling functions. 
Now, we note that, due to the existence of the RG invariant $\kappa$, which relates 
to each other the running parameters $X_z$ and $X$ along the 
RG trajectories 
(Eq.(\ref{ksc16}) in the perturbative regime),
we may trade  
$\tilde{\omega}_{0,1} \left[ \frac{x}{\ell} , X_z , X \right]$ for two 
functions $\omega_{0,1}$ of only $\frac{x}{\ell} $ and $X$.  As a final result, Eq.(\ref{dd.7}) becomes
\beq
{\bf \Sigma}_z [ x , \ell , X_z , X ] = \omega_0  \left[ \frac{x}{\ell} , X_z (x) \right] +\ell^{ 1- g} \: 
\omega_1 \left[ \frac{x}{\ell} , X_z (x)  \right]\:\:\:\: .
\label{dd.8}
\eneq
\noindent
Equation (\ref{dd.8}) provides the leading perturbative approximation 
at weak boundary coupling, as it can be easily 
checked from the explicit formula in Equation (\ref{corr.x2}).
Eq.(\ref{dd.8}) illustrates how the function we explicitly 
use in our calculation can be regarded 
as just an approximation to the exact scaling function for 
${\bf \Sigma}_z ( x )$. A more refined analytical 
treatment might in principle be done by considering 
higher-order contributions in perturbation theory in $S^B_{\bf G}$. 
Alternatively, 
one might resort to a fully numerical approach, similar to the one 
used in Ref.[\onlinecite{sholl}]. 
Yet, due to the absence of an intermediate-coupling phase transition 
in the Kondo effect \cite{hewson}, in our 
opinion resorting to a more sophisticated approach would 
improve the quantitative relation between the 
microscopic ''bare'' system parameters and the ones 
in the effective low-energy long-wavelength model Hamiltonian,
without affecting the main qualitative conclusion 
about the Kondo screening length and its effects. 
 
For this reason, here we prefer to rely on the perturbative RG approach 
extended to the correlation functions which, as we show before, already 
provides reliable and consistent results on the effects 
of the emergence of $\ell_*$ on the physical quantities. 
 
 The obtained estimate of $\xi_K$, although perturbative, provides, via the RG 
relation $k_B T_K=\hbar v_F/\xi_K$, an estimate of the Kondo temperature. When 
the measurements are done at finite temperature, of course thermal effects 
affect the estimate of $\xi_K$: we anyway expect that if the temperature is 
much smaller than $T_K$, then such effects are negligeable. Considering 
that $T_K$ has been estimated of order of tens of $nK$ \cite{buccheri16}, and 
that $T_K$ may be increased by increasing $v_F$, which may be up to hundreds 
$nK$, and by increasing $J'/J$, we therefore expect that with temperatures 
smaller than the bandwidth one can safely extract $\xi_K$. One should 
anyway find a compromise since by increasing $J'/J$ the Kondo length decreases 
(and the Kondo effect itself disappears). A systematic study of thermal 
effects on the estimate of $\xi_K$ is certainly an important subject of 
future work. 
 
\section{Conclusions}
\label{conclusions}

In this paper we have studied the measurement  of the 
Kondo screening length in systems of ultracold atoms in deep optical lattices. 
Our motivation relies primarily on the fact that the detection of the Kondo screening 
length  from experimentally measurable quantities in general appears to be quite a 
challenging task. For this reason, we proposed to perform the measurement in cold 
atom setups, whose parameters can be, in principle, tuned in a controllable way to
desired values. 
 
Specifically, after reviewing  the mapping between  the BH model at half-filling
with inhomogenous hopping amplitudes onto a spin chain Hamiltonian with 
Kondo-like magnetic impurities, we  have proposed 
to extract the Kondo length from a suitable quantity obtained 
by integrating the real space density-density correlation functions. The 
corresponding estimates we recover for the Kondo length are eventually 
found to   assume   values definitely within the reach of present experiments 
($\sim$ tens of lattice sites 
for typical values of the system parameters). We showed that the Kondo length does not significantly depend on nearest-neighbor interaction 
$V$, and it mainly depends on the impurity 
link $J'$.

Concerning the Kondo length,
a comment is in order for quantum-optics oriented readers: in a typical 
measurement of the Kondo effect at a magnetic impurity in a conducting metallic host, 
one has access to the Kondo temperature $T_K$, by just looking at the scale at 
which the resistance (or the conductance, in experiments in quantum dots) bends upwards, on 
lowering $T$. The very existence of the screening length $\xi_K$  is just inferred from the 
emergence of $T_K$ and from the applicability of one-parameter scaling to the Kondo regime, 
which yields   $\xi_K=\hbar v_F/k_B T_K$. However the latter relation stems from the 
validity of the RG approach. Thus,  ultimately 
probing directly $\xi_K$ in solid-state samples would correspond to verifying 
the scaling in the Kondo limit, which is what makes it hard to actually 
perform the measurement. At variance, as we comment 
for solid-state oriented readers, 
in the ultracold gases systems we investigate here, one can certainly 
study dynamics (e.g., tilting the system) but a stationary 
flow of atoms cannot be (so far) established, so that the measure 
of $T_K$ may be an hard task to achieve. Rather surprisingly, as our results 
highlight, it is the Kondo length which can be more easily 
directly detected in ultracold gases and our corresponding estimates  
(order of tens of lattice sites) appear to be rather encouraging in this 
direction.

Several interesting issues deserve in our opinion further work: as first,
it would be desirable to compare the perturbative results we obtain 
 in this paper with numerical, nonperturbative findings in the Bose-Hubbard 
chain, 
to determine the corresponding correction to the value of $\ell_*$. 
  It would be also important to  
understand the corrections to the inferred value of 
$\xi_K$ coming from finite temperature effects, that should be anyway 
negligeable for $T$ (much) smaller that $T_K$. 
Even more importantly, 
we mostly assumed that it is possible to alter the hopping parameters 
in a finite region without affecting the others. This led us to 
infer, for instance, the existence of 
the ensuing even-odd effect -- however, having two lasers 
with $\sigma \ll d$ is a condition that may be straightforwardly implementable.
In this case,  one has to deal with generic space-dependent hopping amplitudes 
$t_{j;j+1}$. It would therefore be of interest to address, very likely within 
a fully numerical approach, the fate of the even-odd effect in the presence of
a   small modulation in space of the outer hopping terms. In particular, 
a  theoretically interesting issue would be the competition 
between an extended nonlocal central region and the occurrence 
of magnetic and/or nonmagnetic impurities in the chain. 
 Another point to be addressed is that an on-site nonuniform potential may in 
 principle be present (event though its  effect may be reduced by hard wall confining 
potentials) and an interesting task is to determine the interplay 
between the Kondo length and the length scale of such an additional  potential.

In conclusion, we believe that our results show that the possible realization of the setup proposed in this 
paper could pave the way to the study of magnetic impurities and, 
in perspective, to the experimental implementation of ultracold realizations 
of Kondo lattices and detection of the Kondo length, 
providing, at the same time, a chance for studying  several 
interesting many-body 
problems in a controllable way.\\

We thank L. Dell'Anna, A. Nersesyan, 
A. Papa, and D. Rossini for valuable discussions. A.T. acknolweldges support 
from talian Ministry of Education and Research (MIUR) Progetto Premiale 2012 ABNANOTECH - ATOM-BASED NANOTECHNOLOGY. 

\appendix

\section{Effective weak link- and Kondo-Hamiltonians for 
a spin-1/2 $XXZ$ spin chain}
\label{kondo_hamiltonians}

In this appendix we review the description of a region {\bf G}, singled out
by weakening two links in a $XXZ$ spin chain, in 
terms of an effective low-energy
Hamiltonian $H_{\bf G}$. In particular, we show how, 
depending on whether the number 
of sites containined within {\bf G} is odd, or even, either $H_{\bf G}$ 
coincides with 
the Kondo Hamiltonian $H_K$ in Eq.(\ref{HAM-K}), or it describes a 
weak link between two 
''half-chains'' \cite{glazman97,giuliano05}. 

In general, Kondo effect in spin-1/2 chains has been studied for an isolated 
magnetic impurity  (the ``Kondo spin''), which may either lie at the end
of the chain (boundary impurity), or at its middle (embedded impurity) 
\cite{eggert92,furusaki98}. In the former case, the impurity can be realized by
``weakening'' one link of the chain, in the latter case, instead, it can be
realized by weakening two links in the body of the chain. 
Following the discussion in Sec.\ref{ren_group} of the main text, 
here we mostly 
focus on the latter case.  In general, in a spin chain, 
impurities may be realized as extended objects, as well, that is, 
as regions containing two, or more, sites. Whether the Kondo physics is
realized, or not, does actually depend on whether the level spectrum of
the isolated impurity takes, or not, a degenerate ground state. A doubly
degenerate ground state is certainly realized in an extended
region with an odd number of sites, without explicit breaking of 
``spin inversion'' symmetry (that is, in the absence of local ``magnetic
fields''). For instance, let us consider a central region realized by
three sites ($j=-1,0,1$), lying between the weak links. Let the central region
Hamiltonian be given by 
\beq
H_{3J}^{\rm Middle} = - J \left( S_{-1}^+ S_0^- + S_0^+ S_1^- + {\rm h.c.} \right)
+ J^z \left( S_{-1}^z S_0^z + S_0^z S_1^z \right)
\:\:\:\: , 
\label{eo2}
\eneq
\noindent
and let the central region be connected to the left-hand chain 
(which, as in the 
main text, we denote by the label $_L$), 
and to the right-hand chain (denoted by 
the label $_R$) with the coupling Hamiltonian
\beq
H_{\rm Coupling} = - \left( 
J_L^{'} S_{1,L}^+S_{-1}^- + J_R^{'} S_{1,R}^+S_{1}^- + {\rm h.c.} \right)
+  \left( J_{z , L} ^{'}S_{1,L}^zS_{-1}^z + J_{z , R} ^{'}S_{1,R}^z S_{1}^z \right)
\:\:\:\: . 
\label{eo12}
\eneq
\noindent

A simple algebraic calculation shows that the ground state of 
$H_{3J}^{\rm Middle}$ is doubly degenerate and consists of the spin-1/2
doublet given by
\beq
| \frac{1}{2} \rangle_2 = \frac{1}{\sqrt{2}} \{ \sin ( \frac{\theta}{2} )
[ \uparrow \uparrow 
\downarrow \rangle + | \downarrow \uparrow \uparrow \rangle ] 
+ \sqrt{2} \cos ( \frac{\theta}{2} ) | \uparrow \downarrow \uparrow \rangle \}
\;\;\;\; , 
\label{eo6}
\eneq
\noindent
and
\beq
| - \frac{1}{2} \rangle_2 = \frac{1}{\sqrt{2}} \{ \sin ( \frac{\theta}{2} )
[ \downarrow \downarrow 
\uparrow \rangle + | \uparrow \downarrow \downarrow \rangle ] 
+ \sqrt{2} \cos ( \frac{\theta}{2} ) | \downarrow \uparrow \downarrow \rangle \}
\;\;\;\; , 
\label{eo7}
\eneq
\noindent
with
\beq
\cos ( \theta ) = \frac{J_z}{\sqrt{2 J^2 + J_z}} \;\;\; , \;\;
\cos ( \theta ) = \frac{\sqrt{2} J}{\sqrt{2 J^2 + J_z}}
\:\:\:\: , 
\label{eo8}
\eneq
\noindent
whose energy is given by $E_2^{\frac{1}{2}} = - J_z - \sqrt{J_z^2 + 2 J^2}$.
Defining an effective spin-1/2 operator for the central region, ${\bf S}_G$, as
\beq
S_G^+ \equiv |  \frac{1}{2} \rangle_2 ~_2 \langle  - \frac{1}{2} | 
\:\:\: , \:\:
S_G^z \equiv \frac{1}{2} 
\sum_{ b = \pm 1} b |  b \frac{1}{2} \rangle_2 ~_2 \langle  b \frac{1}{2} | 
\:\:\:\: , 
\label{eo13}
\eneq
\noindent
allows to rewrite $H_{3J}^{\rm Middle} + H_{\rm Coupling}$ as
\beq
V_{B}^{\rm 3J} = -\{ [  J^{'}_L \sin ( \theta ) S_{1 , L}^+ + J^{'}_R  S_{1 , R}^+ ] S_G^- 
+  [ J^{'}_L  S_{1 , L}^- + J^{'}_R S_{1 ,R}^- ] S_G^+ \} + \cos ( \theta ) 
[ J_{z , L}^{'}     S_{1 , L}^z +J_{z , R}^{'}  S_{1 , R}^z ] S_G^z 
\:\:\:\: . 
\label{eo14}
\eneq
\noindent
Thus, we see that we got back to the spin-1/2 spin-chain Kondo 
Hamiltonian, with a renormalization
of the boundary couplings, according to
\beq
J^{'}_{L (R)} \longrightarrow J^{'}_{L (R)} \sin ( \theta ) = \frac{ \sqrt{2} J^{'}_{L (R)} J}{ \sqrt{
2 J^2 + J_z}} \;\;\; , \;\;
 J_{z , L(R)}^{'} \longrightarrow J_{z , L(R)}^{'}  \cos ( \theta ) = 
\frac{  J_{z , L(R)}^{'}J_z}{ \sqrt{
2 J^2 + J_z}}
\:\:\:\: . 
\label{eo15}
\eneq
\noindent
A local magnetic field $h$ may break the ground state degeneracy, thus 
leading, in principle, to the breakdown of the Kondo effect. However, in
analogy to what happens in a Kondo dot in the presence of an external
magnetic field \cite{gg98_1,gg98_2,gg98_3}, 
Kondo physics should survive, at least
as long as $h \ll {\cal E}_K$, with ${\cal E}_K (\sim k_B T_K)$ being the typical energy 
scale associated to the onset of Kondo physics.

At variance, when the central region is made by an even number of
sites, the groundstate is not degenerate anymore. As a consequence, the
central region should be regarded as a weak link between two chains. For instance, we may
consider the  case in which the central region is made
by two sites. Using for the various parameters the same symbols we used above, 
performing a SW resummation, we obtain the effective weak link 
boundary Hamiltonian
\beq
V_{B}^{\rm 2J} = - \lambda_\perp \left( S_{L , 1}^+ S_{R , 1}^- + 
S_{R , 1}^+ S_{L , 1}^-
\right) -  \lambda_z S_{ L , 1}^z S_{R , 1}^z 
\:\:\:\: , 
\label{eoc21}
\eneq
\noindent
with
\beq
\lambda_\perp \sim \frac{ ( J^{'} )^2}{ J + 2 J_z} \;\;\; , \;\;
\lambda_z \sim \frac{ ( J_z^{'} )^2}{ 2 J}
\:\:\:\: . 
\label{eoc22}
\eneq
\noindent

\section{bosonization approach to impurities in the $XXZ$ spin chain}
\label{boso_imp}

In this section we review the bosonization approach to the $XXZ$ spin chain
as it was originally developed in Refs.[\onlinecite{eggert92,furusaki98}]. 
As a starting point, we consider a single, homogeneous spin-1/2 $XXZ$ spin chain,
with $\ell$ sites, obeying  open boundary conditions at its endpoints,
described by the model Hamiltonian $H_{\rm XXZ}$, given by   
\beq
H_{\rm XXZ} = - J \sum_{ j = 1}^{\ell-1} \left( S_j^+ S_{j+1}^- + S_{j+1}^+ 
S_j^- \right) + J^z \sum_{ j = 1}^{\ell -1} S_j^z S_{j+1}^z 
\:\:\:\: . 
\label{core1}
\eneq
\noindent
The low-energy, long-wavelength dynamics of such a chain is 
described \cite{eggert92} in terms of a spinless, real bosonic
field $\Phi ( x , \tau )$ and of its dual field $\Theta ( x , \tau )$.
The imaginary time action for $\Phi$ is given by
\beq
S_E [ \Phi ] = \frac{g}{ 4 \pi} \: \int_0^\beta \: d \tau \: \int_0^\ell \: d x \:
\left[ \frac{1}{ u } \left( \frac{ \partial \Phi}{ \partial \tau} \right)^2
+ u \left( \frac{ \partial \Phi}{ \partial x} \right)^2 \right]
\:\:\:\: , 
\label{core2}
\eneq
\noindent
where the constants $g,u$ are given by
\beq
g = \frac{\pi}{2 ( \pi - {\rm arccos} ( \frac{\Delta}{2} ))} 
\;\;\; , \;\;
u = v_f \left[ \frac{\pi}{2} \frac{\sqrt{1 - (\frac{\Delta}{2})^2}}{ 
{\rm arccos ( \frac{\Delta}{2} )}
} \right]
\;\;\;\; , 
\label{core3}
\eneq
\noindent
with $v_f = 2 d J$, $d$ being the lattice step, and $\Delta = J^z / J$. 
The fields $\Phi$ and $\Theta$
are related to each other by the relations $\frac{\partial \Phi ( x , \tau )}{
\partial x} =  \frac{1}{u} \frac{\partial \Theta ( x , \tau )}{ \partial x}$,
and $\frac{\partial \Theta ( x , \tau )}{
\partial x} =  \frac{1}{u} \frac{\partial \Phi ( x , \tau )}{ \partial x}$.
A careful bosonization procedure shows that, in addition to the free
Hamiltonian in Eq.(\ref{core2}), an additional Sine-Gordon, Umklapp interaction
arises, given by
\beq
H_{\rm L}^{\rm SG} = - G_U \int_0^\ell \: d x \:  \cos [ 2 \sqrt{2} 
\Theta ( x ) ]  
\:\:\:\: . 
\label{ksc3}
\eneq
\noindent
Since the scaling dimension of $H_{\rm L}^{\rm SG}$ is $h_U = 4 g $, it
will be always irrelevant within the window of values of $g$ we are 
considering here, that is, $1/2 < g $. 
In fact, $H_{\rm L}^{\rm SG}$ becomes
marginally irrelevant at the ``Heisenberg point'', $g=1/2$, 
which deserves special attention \cite{sorensen}, 
though we do not consider it here.  Within the continuous bosonic field
framework, the open boundary conditions of
the chain are  accounted for by imposing Neumann-like boundary
conditions on the field $\Phi ( x , \tau)$ at both boundaries \cite{giuliano05,giuliano07,giuliano09,cmgs}, 
that is

\beq
\frac{ \partial \Phi ( 0 , \tau )}{ \partial x } = \frac{ \partial
\Phi ( \ell , \tau )}{ \partial x} = 0 
\:\:\:\: . 
\label{neu1}
\eneq
\noindent
Equation (\ref{neu1}) implies the following mode expansions
for $ \Phi ( x , \tau )$ and $\Theta ( x , \tau )$ 
\begin{eqnarray} 
\Phi ( x , \tau ) = \sqrt{\frac{2}{g}} \left\{ q - \frac{i \pi u \tau }{\ell} P
+ i \sum_{ n \neq 0} \frac{ \alpha ( n )}{ n} \cos \left[ \frac{ \pi n x}{\ell} 
\right] e^{ - \frac{\pi n }{\ell} u \tau} \right\}
\nonumber \\
\Theta ( x , \tau ) = \sqrt{2g} \left\{ \theta + \frac{ \pi x }{\ell} P
+  \sum_{ n \neq 0} \frac{ \alpha ( n )}{ n} \sin \left[ \frac{ \pi n x}{\ell} 
\right] e^{ - \frac{\pi n }{\ell} u \tau} \right\}
\:\:\:\: ,
\label{core12}
\end{eqnarray}
\noindent
with the normal modes satisfying the algebra
\beq
[ q , P ] = i \;\;\; , \;\; 
[ \alpha ( n ) , \alpha (n' ) ] = n \delta_{ n + n' , 0 }
\:\:\:\: . 
\label{cc.12}
\eneq
\noindent
The bosonization procedure allows for expressing the spin operators 
in terms of the $\Phi$- and $\Theta$-fields. The result is \cite{hikifu_1}  
\begin{eqnarray} 
S_j^+ &\longrightarrow &    \left\{ c (-1)^j
 e^{ \frac{i}{ \sqrt{2}} \Phi ( x_j , \tau ) } + b 
 e^{  
\frac{i}{ \sqrt{2}} \Phi ( x_j , \tau )  + i \sqrt{2} \Theta ( x_j , \tau )}
\right\}
\nonumber \\
S_j^z & \longrightarrow & \left[  \frac{1}{\sqrt{2} \pi} \frac{ \partial 
\Theta ( x_j , \tau )}{ \partial x} + a  (-1)^j 
\sin [ \sqrt{2} \Theta ( x_j  , \tau ) ] \right]
\:\:\:\: . 
\label{core3bis}
\end{eqnarray}
\noindent
The numerical parameters $a,b,c$ in Eq.(\ref{core3bis}) depend only on  the 
anisotropy parameter $\Delta = J_z / J$ 
\cite{hikifu_1,hikifu_2,shashi,lky_1,lky_2}.
While their actual values is not essential to the RG
analysis in Sec.\ref{ren_group}, 
it becomes important when computing the real-space 
correlation functions of the chain within the bosonization approach, in which 
case one may refer to the extensive literature on the subject, 
as we do in Sec.\ref{density_density}.

To employ the bosonization approach to  study an impurity created between 
the $_L$ and the $_R$ chain, we start by doubling the  construction
outlined above, so to separately bosonize the two chains with 
open boundary conditions
(which is appropriate in the limit of a weak interaction strength for either 
$H_K$ in Eq.(\ref{HAM-K}), or $V_{B}^{\rm 2J}$ in Eq.(\ref{eoc21})). Therefore, 
on introducing two pairs of conjugate bosonic fields $\Phi_L , \Theta_L$ and 
$\Phi_R , \Theta_R$ to describe the two chains, the corresponding 
Euclidean action is given by 
\beq
S_E [ \Phi_L , \Phi_R ] = \frac{g}{ 4 \pi} \: \int_0^\beta \: d \tau \: \int_0^\ell \: d x \:
\sum_{ X = L , R} \: \left[ \frac{1}{ u } \left( \frac{ \partial \Phi_X}{ \partial \tau} \right)^2
+ u \left( \frac{ \partial \Phi_X}{ \partial x} \right)^2 \right]
\:\:\:\: ,
\label{corex}
\eneq
\noindent
supplemented with the boundary conditions 
\beq
\frac{\partial \Phi_L ( x , 0 ) }{\partial x} = \frac{\partial \Phi_L ( \ell , \tau ) }{
\partial x } = 0 
\;\;\; ,\;\;
\frac{\partial \Phi_R ( x , 0 ) }{\partial x} = \frac{\partial \Phi_R ( \ell , \tau ) }{
\partial x } = 0
\:\:\:\: .
\label{bcx}
\eneq
\noindent
Taking into account the bosonization recipe for the spin-1/2 operators, 
Eqs.(\ref{core3bis}), one obtains that, in
the case in which {\bf G} contains an even number of sites (and is, therefore, 
described by the prototypical impurity Hamiltonian $V_{B}^{\rm 2J}$),  
the effective weak link impurity between the two chains 
is described by the Euclidean action
\beq
S_{\bf G}^{B}
 = - \lambda_\perp \: \int_0^\beta \: d \tau \: \{ e^{ \frac{i}{\sqrt{2}}
[ \Phi_L ( \tau ) - \Phi_R ( \tau )]} + e^{ - \frac{i}{\sqrt{2}}
[ \Phi_L ( \tau ) - \Phi_R ( \tau )]} \} -
\frac{ \lambda_z}{ 2 \pi^2}  \: \int_0^\beta \: d \tau \: \frac{ \partial
\Theta_L ( \tau )}{ \partial x} \frac{ \partial
\Theta_R ( \tau )}{ \partial x}  
\;\;\;\; , 
\label{bbi2}
\eneq
\noindent
with  $\Phi_{L , R} ( \tau ) \equiv \Phi_{L , R}  ( 0 , 
\tau)$, and $\Theta_{L , R}  ( \tau ) \equiv \Theta_{ L , R}  ( 0 , \tau )$. 
Similarly, 
in the case in which {\bf G} contains an odd number of sites, 
in bosonic coordinates, the
prototypical Kondo Hamiltonian $H_K$ yields to the Euclidean action
given by   
\beq
S_{\bf G}^{B} = -   \int_0^\beta \: d \tau \: \{ [ J^{'}_L e^{ \frac{i}{\sqrt{2}}
 \Phi_L ( \tau ) } + J^{'}_R  e^{ \frac{i}{\sqrt{2}} \Phi_R ( \tau ) } ] S_{\bf G}^-
+ {\rm h.c.} \}
+ \frac{1}{ \sqrt{2} \pi}  \int_0^\beta \: d \tau \: \left\{ \left[ 
J_{z,L}^{'} \frac{ \partial \Theta_L ( \tau )}{ \partial x} + 
J_{z,R}^{'} \frac{ \partial \Theta_R ( \tau )}{ \partial x} \right] S_{\bf G}^z \right\}
\:\:\:\: . 
\label{bbi3}
\eneq
\noindent
Equation (\ref{bbi3}) provides the starting point to 
perform the RG analysis for the Kondo impurity of Section
\ref{ren_group}. To illustrate in detail the application of the 
RG approach to link impurities in spin chains, in the following 
part of this appendix we employ it to study the  weak link boundary 
action in Eq.(\ref{bbi2}). Following the standard RG recipe, to  
describe how the relative weight of the impurity 
interaction depends on the reference cutoff scale of the system, 
we have  to recover the corresponding RG 
scaling equations for the running coupling strengths associated to 
$\lambda_z$ and to $\lambda_\perp$. This is readily done by resorting 
to the Abelian bosonization approach to 
spin chains applied to the boundary action in Eq.(\ref{bbi2}) \cite{eggert92}. 
From  Eq.(\ref{bbi2}) one readily recovers  the 
scaling dimensions of the various terms 
from standard Luttinger liquid techniques, once one has assumed the 
mode expansions in Eqs.(\ref{core12}) for the fields $\Phi_L ( x , \tau )  , \Theta_L ( x , \tau )$, as 
well as $\Phi_R (  x , \tau ) , \Theta_R ( x , \tau )$ \cite{eggert92,furusaki98}. 
Specifically, one finds that the term 
$\propto \lambda_\perp$ has scaling dimension 
$h_\perp = \frac{1}{g}$, while the term 
$\propto \lambda_z$ has scaling dimension $h_\parallel = 2$. 
As we use the chain length $\ell$ as 
scaling parameter of the system, to keep in touch with the 
standard RG approach, we define the dimensionless 
running coupling strengths ${\cal L}_\perp ( \ell ) = \left( \frac{\ell}{\ell_0} \right)^{ 1 - \frac{1}{g} }  \frac{\lambda_\perp}{J} $ and 
${\cal L}_\parallel ( \ell ) = \left( \frac{\ell}{\ell_0} \right)^{ -1 } \frac{\lambda_z}{J} $, with 
$\ell_0$ being a reference length scale 
(see below for the discussion on the estimate of $\ell_0$).  
To leading order in the coupling strengths, we obtain the 
perturbative RG  equations for the running parameters 
given by \cite{kane92} 
\begin{eqnarray}
 \frac{d {\cal L}_\perp ( \ell ) }{d \ln \left( \frac{\ell}{\ell_0} \right)} &=& \left[ 1 - \frac{1}{g} \right] {\cal L}_\perp ( \ell ) \nonumber \\
  \frac{d {\cal L}_\parallel  ( \ell ) }{d \ln \left( \frac{\ell}{\ell_0} \right)} &=& -  {\cal L}_\parallel  ( \ell )
  \:\:\:\: . 
  \label{kk.1} 
\end{eqnarray}
\noindent
Equations (\ref{kk.1}) encode the main result concerning the dynamics of 
a weak link in an otherwise uniform $XXZ$ chain \cite{eggert92,eggert99,giuliano05}. 
Leaving aside the trivial case $g=1$, corresponding to 
effectively noninteracting JW fermions, which 
do not induce any universal 
(i.e., independent of the bare values of the system parameters) 
flow towards a conformal fixed point, we see that the behavior of the 
running strengths on 
increasing $\ell$ is drastically different, according to whether 
$g<1 \; (\Delta >0)$, or 
$g>1 \; (\Delta < 0 )$. In the former case, 
both $h_\perp$ and $h_\parallel$ are $>1$, which 
implies that  $V_B^{2J}$ is an irrelevant perturbation to the disconnected 
fixed point. The impurity interaction strengths flow to zero in the low-energy,
long-wavelength limit, that is, under RG  trajectories, 
the  system flows back towards  the fixed point corresponding 
to two disconnected chains. At variance, when 
$g>1$, ${\cal L}_\perp ( \ell )$ grows along the RG trajectories and  the system   flows 
towards a ''strongly coupled'' fixed point, which 
corresponds to the healed chain, in which the weak link has been 
healed within an effectively uniform chain obtained 
by merging the two side chains with each other 
\cite{kane92}. The healing takes place at a scale 
$\ell \sim \ell_{\rm Heal}$, with \cite{glazman97,giuliano05}
\beq
\ell_{\rm Heal} \sim \ell_0 \left( \frac{1}{{\cal L} ( \ell_0 ) } \right)^\frac{g}{g-1}
\:\:\:\: . 
\label{healing}
\eneq
\noindent
As we see from Eq.(\ref{healing}), defining $\ell_{\rm Heal}$ 
requires introducing a nonuniversal, reference length scale 
$\ell_0$. $\ell_0$ is (the plasmon velocity times) the reciprocal 
of the high-energy cutoff $D_0$ of our system. 
To estimate $D_0$, we may simply require that 
we cutoff all the processes at energies at which the approximations 
we employed in Appendix \ref{boso_imp} to get 
the effective boundary Hamiltonians break down. This means that $D_0$ must be 
of the order of the energy difference $\delta E$ between the 
groundstate(s) and the first excited state
of the central region Hamiltonian. From the discussion of 
Appendix \ref{boso_imp}, we see that
  $\delta E \sim J$, which, since we normalized all the 
running couplings to $J$, implies 
$\ell_0 \sim d$, $d$ being the lattice step of the 
microscopic lattice Hamiltonian describing 
our spin system. To conclude, it is important to 
stress   that, though an RG invariant 
length scale $\ell_{\rm Heal}$ emerges already at a weak link between 
two chains with $\Delta < 0$, there is no  screening cloud 
associated to this specific problem. 
Indeed, in  the case of a weak link impurity, 
the healing of the chain is merely a consequence of repeated scattering 
off the Friedel oscillations due to backscattering at the weak link
\cite{matvevgla_1,matvegla_2,giu_nava},
which conspire to fully heal the impurity at 
a scale $\ell_{\rm Heal}$. 
At variance, when there is 
an active spin-1/2 impurity, 
the density oscillations are no longer simply determined by
the scattering by Friedel oscillations, but there is also the emergence of 
the Kondo screening cloud induced in the system \cite{eggert99}.  
 
\section{Bosonization results for the correlation functions between spin operators at 
finite imaginary time}
\label{cor_functions}

In this appendix we provide the generalization of the equal-time 
spin-spin correlation functions on an open chain, derived in 
Ref.[\onlinecite{hikifu_1}], to the case in which the spin operators 
are computed at different imaginary times $\tau , \tau'$. As discussed in 
the main text, such a generalization is a necessary step in order to 
compute the contributions to the spin correlations due to the impurity 
interaction in $S_{\bf G}^B$. The starting point is provided by the 
finite-$\tau$ bosonic operators over a homogeneous, 
finite-size chain of length $\ell$, 
which we provide in Eqs.(\ref{core12}) of the main text. Inserting those
formulas for $\Phi ( x , \tau ) $ and $\Theta ( x , \tau )$ in the bosonic 
formulas in Eqs.(\ref{core3bis}) and computing the imaginary-time ordered 
correlation functions $G_{+ , -}  ( x , x'  ; \tau | \ell ) = 
\langle {\bf T}_\tau S^+_x ( \tau ) S_{x'}^- ( 0 ) \rangle $ and 
$G_{z , z}  ( x , x'  ; \tau | \ell ) = 
\langle {\bf T}_\tau S^z_x ( \tau ) S_{x'}^z ( 0 ) \rangle $, one obtains:
\begin{eqnarray}
&& G^{+-} ( x , x' ; \tau  | \ell ) = \nonumber \\
&& c^2 (-1)^{ x - x'} \left| \frac{2\ell}{ \pi} 
\sin \left( \frac{\pi x}{\ell} \right)
\right|^{\frac{1}{4 g }} \left| \frac{2\ell}{ \pi} \sin \left( 
\frac{\pi x' }{\ell} \right) \right|^{\frac{1}{4 g }}  \left| \frac{2\ell}{ \pi} \sinh \left( 
\frac{\pi}{2\ell} [ u \tau  
+ i ( x - x' ) ] \right) \right|^{ - \frac{1}{2g}} 
\left| \frac{2\ell}{\pi} \sinh \left( \frac{\pi}{2\ell} [ u  \tau 
+ i ( x + x' ) ] \right) \right|^{ - \frac{1}{2g}} 
\nonumber \\
&& + b^2 \left| \frac{2\ell}{ \pi} \sin \left( \frac{\pi x}{\ell} \right)
\right|^{\frac{1}{4 g }-g} \left| \frac{2\ell}{ \pi} \sin \left( 
\frac{\pi x' }{\ell} \right) \right|^{\frac{1}{4 g }-g} 
 \left| \frac{2\ell}{\pi} \sinh \left( \frac{\pi}{2\ell} [ u  \tau  
+ i ( x - x' ) ] \right) \right|^{ - \frac{1}{2g}-2g} 
\left| \frac{2\ell}{ 
\pi} \sinh \left( \frac{\pi}{2\ell} [ u \tau  
+ i ( x + x' ) ] \right) \right|^{ - \frac{1}{2g}+2g} 
\nonumber \\
&& + bc \;  {\rm sgn} ( x - x' ) \left| \frac{2\ell}{ \pi}
 \sin \left( \frac{\pi x}{\ell} \right)
\right|^{\frac{1}{4 g }} \left| \frac{2\ell}{ \pi} \sin \left( 
\frac{\pi x' }{\ell} \right) \right|^{\frac{1}{4 g }} 
\left| \frac{2\ell}{ 
\pi} \sinh \left( \frac{\pi}{2\ell} [ u  \tau  
+ i ( x - x' ) ] \right) \right|^{ - \frac{1}{2g}} 
\left| \frac{2\ell}{ 
\pi} \sinh \left( \frac{\pi}{2\ell} [ u  \tau  
+ i ( x + x' ) ] \right) \right|^{ - \frac{1}{2g}} 
\nonumber \\
&& \times \left[ (-1)^x \left| \frac{2\ell}{ 
\pi} \sin \left( \frac{\pi x'}{\ell}
\right) \right|^{-g} -  (-1)^{x'} 
\left| \frac{2\ell}{ \pi} \sin \left( \frac{\pi x}{\ell} \right) \right|^{-g}
\right]
\:\:\:\: ,
\label{corr.x1}
\end{eqnarray}
\noindent
as well as 
\begin{eqnarray}
&& G^{zz} ( x , x' ; \tau | \ell  )  = 
- \frac{g}{4 \ell^2} \biggl\{ \left[ \frac{1 - \cosh \left(   \frac{\pi u  \tau  
}{\ell }  \right)\cos \left( \frac{\pi  ( x - x' )}{\ell} 
\right) }{ \left[ 1 + \cos^2 \left( \frac{\pi  ( x - x' )}{\ell} 
\right)  - 2 \cos \left( \frac{\pi  ( x - x' )}{\ell} 
\right)  \cosh \left(   \frac{\pi u  \tau  
}{\ell }  \right) + \sinh^2 \left(   \frac{\pi u  \tau  
}{\ell }  \right)\right]}  \right] \nonumber \\
&& + \left[ \frac{1 - \cosh \left(   \frac{\pi u  \tau  
}{\ell }  \right)\cos \left( \frac{\pi  ( x + x' )}{\ell} 
\right) }{ \left[ 1 + \cos^2 \left( \frac{\pi  ( x + x' )}{\ell} 
\right)   - 2 \cos \left( \frac{\pi  ( x + x' )}{\ell} 
\right)  \cosh \left(   \frac{\pi u  \tau  
}{\ell }  \right) + \sinh^2 \left(   \frac{\pi u  \tau  
}{\ell }  \right)\right]}  \right]  \biggr\}
\nonumber \\
&& + \frac{a^2}{2} (-1)^{x- x'}  
\left| \frac{2\ell}{\pi} \sin \left( \frac{\pi x}{\ell} \right)
\right|^{-g} \left| \frac{2\ell}{\pi} \sin \left( \frac{\pi x'}{\ell} \right)
\right|^{-g}  \times
\nonumber \\
&& \biggl\{ 
\left| \frac{ \sinh \left( \frac{\pi}{2\ell} [ u  \tau + i 
( x - x' ) ] \right)}{ \sinh \left( \frac{\pi}{2\ell} [ 
u  \tau  + i ( x + x' ) ] \right) } \right|^{-2g} 
- \left| \frac{ \sinh \left( \frac{\pi}{2\ell} [ u  \tau + i 
( x - x' ) ] \right)}{ \sinh \left( \frac{\pi}{2\ell} [ 
u \tau  + i ( x + x' ) ] \right) } \right|^{2g} 
\biggr\}
\nonumber \\
&& - \frac{a ig}{2 \ell} (-1)^{x'}
\left| \frac{2 \ell}{\pi} \sin \left( \frac{\pi x' }{\ell} \right)
\right|^{-g}   \times
\nonumber \\
&& \biggl\{ \coth \left[ \frac{\pi}{2\ell} ( u  \tau  
+ i ( x + x' ) ) \right] - 
\coth \left[ \frac{\pi}{2\ell} ( u  \tau  
- i ( x + x' ) ) \right] 
 - \coth \left[ \frac{\pi}{2\ell} ( u  \tau 
+ i ( x - x' ) ) \right] + 
\coth \left[ \frac{\pi}{2\ell} ( u \tau  
- i ( x - x' ) ) \right]  \biggr\}
\nonumber \\
&& - \frac{a ig}{2 \ell} (-1)^x
\left| \frac{2\ell}{\pi} \sin \left( \frac{\pi x}{\ell} \right) 
 \right|^{-g} \times
\nonumber \\
&& \biggl\{ \coth \left[ \frac{\pi}{2\ell} ( u  \tau 
+ i ( x + x' ) ) \right] - 
\coth \left[ \frac{\pi}{2\ell} ( u  \tau 
- i ( x + x' ) ) \right] 
+ \coth \left[ \frac{\pi}{2\ell} ( u  \tau  
+ i ( x - x' ) ) \right] -
\coth \left[ \frac{\pi}{2\ell} ( u  \tau  
- i ( x - x' ) ) \right]  \biggr\} \:\:\:\: . \nonumber \\
\label{corr.x2}
\end{eqnarray}
\noindent
As stated above, Eqs.(\ref{corr.x1}) and (\ref{corr.x2}) provide  the 
finite-$\tau$ generalization of Eqs.(8a) and (8b) of Ref.[\onlinecite{hikifu_1}],
to which they reduce in the $\tau \to 0$ limit.
 
\bibliography{kca_revised_V3.bib}
\end{document}